\documentclass[longauth]{aa}
\usepackage[varg]{txfonts}
\usepackage{natbib}
\usepackage{graphicx}
\bibpunct{(}{)}{;}{a}{}{,}
\newcommand{\parhead}[1]{\paragraph{\emph{#1:}}}
\usepackage{hyperref}
\makeatletter
\renewcommand*\aa@pageof{, page \thepage{} of \pageref*{LastPage}}
\makeatother

\begin{document}

\title{Complex Organic Molecules in Protostars with ALMA Spectral Surveys (COMPASS)}
\subtitle{I. Overview of the ALMA Large Program}
\author{J. K. J{\o}rgensen\inst{\ref{inst1}} \and A. Coutens\inst{\ref{inst2}} \and M. N. Drozdovskaya\inst{\ref{inst3},\ref{inst3b}} \and J.-E. Lee\inst{\ref{inst4}} \and A. L. Plunkett\inst{\ref{inst5}} \and S. Maret\inst{\ref{inst6}} \and A. Belloche\inst{\ref{inst7}} \and D. Harsono\inst{\ref{inst8}} \and Z. Telkamp\inst{\ref{inst9}} \and F. Cruz-S\'aenz de Miera\inst{\ref{inst2}} \and M. L. R. van 't Hoff\inst{\ref{inst10}} \and \'A. K\'osp\'al\inst{\ref{inst11},\ref{inst12},\ref{inst13}} \and M. Rao\inst{\ref{inst9}} \and J. Bergner\inst{\ref{inst14}} \and J.-H. Jeong\inst{\ref{inst4}} \and S. Spezzano\inst{\ref{inst15}} \and B. A. McGuire\inst{\ref{inst16},\ref{inst5}} \and P. Nazari\inst{\ref{inst17}} \and Y.-L. Yang\inst{\ref{inst19}} \and H.-S. Yun\inst{\ref{inst4},\ref{inst18}} \and A. Andreu\inst{\ref{inst2}} \and J. Ferrer Asensio\inst{\ref{inst19}} \and C.-H. Kim\inst{\ref{inst4}}  \and N. F. W. Ligterink\inst{\ref{inst20}} \and Y. Lin\inst{\ref{inst15}} \and S.-Y. Liu\inst{\ref{inst21}} \and M. L{\"u}tzen\inst{\ref{inst1}} \and P. Marchand\inst{\ref{inst2}} \and C. Xue\inst{\ref{inst16}} \and S. Zeng\inst{\ref{inst19}}}
\institute{Niels Bohr Institute, University of Copenhagen, Jagtvej 155A, DK~2200 Copenhagen N, Denmark, \email{jeskj@nbi.ku.dk}\label{inst1}
\and
Institut de Recherche en Astrophysique et Plan\'etologie, Universit\'e de Toulouse, CNRS, CNES, 9 av. du Colonel Roche, 31028 Toulouse Cedex 4, France\label{inst2}
\and
Physikalisch-Meteorologisches Observatorium Davos und Weltstrahlungszentrum (PMOD/WRC), Dorfstrasse 33, 7260 Davos Dorf, Switzerland\label{inst3}
\and
Department of Chemistry, Biochemistry and Pharmaceutical Sciences (DCBP), Universit{\"a}t Bern, Switzerland \label{inst3b}
\and
Department of Physics and Astronomy, Seoul National University, 1 Gwanak-ro, Gwanak-gu, Seoul 08826, Republic of Korea\label{inst4}
\and National Radio Astronomy Observatory, 520 Edgemont Road, Charlottesville, VA 22903, USA\label{inst5}
\and Univ. Grenoble Alpes, CNRS, IPAG, 38000 Grenoble, France\label{inst6}
\and Max-Planck-Institut f{\"u}r Radioastronomie, Auf dem H{\"u}gel 69, 53121 Bonn, Germany\label{inst7}
\and Institute of Astronomy, Department of Physics, National Tsinghua University, No. 101, Section 2, Kuang-Fu Road, Hsinchu 30013, Taiwan, R.O.C\label{inst8}
\and Department of Astronomy, University of Virginia, 530 McCormick Road, Charlottesville, VA 22904, USA\label{inst9}
\and Purdue University, 525 Northwestern Ave, West Lafayette, IN 47907, USA\label{inst10}
\and Konkoly Observatory, HUN-REN Research Centre for Astronomy and Earth Sciences, MTA Centre of Excellence, Konkoly-Thege Mikl\'os \'ut 15-17, 1121 Budapest, Hungary\label{inst11}
\and Institute of Physics and Astronomy, ELTE E\"otv\"os Lor\'and University, P\'azm\'any P\'eter s\'et\'any 1/A, 1117 Budapest, Hungary\label{inst12}
\and Max-Planck-Insitut f\"ur Astronomie, K\"onigstuhl 17, 69117 Heidelberg, Germany\label{inst13}
\and UC Berkeley, Department of Chemistry, Berkeley, CA 94720, USA\label{inst14}
\and Max-Planck-Institut f{\"u}r extraterrestrische Physik, Giessenbachstrasse 1, 85748 Garching bei M{\"u}nchen, Germany\label{inst15}
\and Department of Chemistry, Massachusetts Institute of Technology and National Radio Astronomy Observatory, 77 Massachusetts Ave, Cambridge, MA 02139 USA; 520 Edgemont Road, Charlottesville, VA 22903, USA\label{inst16}
\and European Southern Observatory, Karl-Schwarzschild-Strasse 2, 85748 Garching, Germany\label{inst17}
\and Star and Planet Formation Laboratory, RIKEN Pioneering Research Institute, Wako-shi, Saitama, 351-0198, Japan\label{inst19}
\and Korea Astronomy and Space Science Institute, 776 Daedeok-daero, Yuseong-gu, Daejeon 34055, Republic of Korea\label{inst18}
\and Faculty of Aerospace Engineering, Delft University of Technology, Kluyverweg 1, 2629 HS Delft, The Netherlands\label{inst20}
\and Institute of Astronomy and Astrophysics, Academia Sinica, 11F of AS/NTU Astronomy-Mathematics Building, No.1, Sec. 4, Roosevelt Rd, Taipei 10617, Taiwan, R.O.C.\label{inst21}
}
\date{Received <date> / Accepted <date>}

\abstract 
{It remains a fundamental research question in astrochemistry to characterize the inventory of complex organic molecules formed during the early stages of star formation, which may affect the eventual chemical composition of protoplanetary disks and potentially planets. So far, only studies of individual sources or selected samples of molecular species have been conducted and the relative importance of environmental and evolutionary effects on the chemical composition remains unclear.}
{The Atacama Large Millimeter/submillimeter Array (ALMA) provides the angular resolution and sensitivity to zoom in on the hot ($T > 100$~K) gas surrounding embedded solar-type protostars and chart their molecular inventories with high accuracy. The ALMA Large Program Complex Organic Molecules in Protostars with ALMA Spectral Surveys (COMPASS) aims to provide comprehensive inventories of the molecular content of a sample of protostellar sources to understand the chemical impact of their environments and evolutionary stages.}
{COMPASS is an unbiased spectral survey of 11 line-rich Class 0 and I protostellar sources in the spectral range from 279.0 to 311.7~GHz at 0.15--0.5$''$ angular resolution with ALMA, corresponding to the inner $\sim$100~au (radius) around the targeted sources. Through this survey, unique chemical fingerprints of each source can be established.}
{This paper provides an overview of the COMPASS program and the adopted strategy in terms of observational setups and targeted sources. We also present a qualitative comparison of the spatial distributions of selected molecules and the relative strengths of the line fluxes of selected complex organics.}
{The qualitative comparisons suggest that chemical differences between sources may be present, for example, between the abundances of groups of oxygen- and nitrogen-bearing species, but also that such variations occur at levels smaller than an order of magnitude. Quantitative comparisons of abundances at these levels require the large bandwidth of the survey, as well as a careful analysis of the excitation and strengths of the several thousands of lines detected toward each source in the survey, as will be explored in forthcoming papers.}

\titlerunning{Overview of the COMPASS ALMA Large Program}
\authorrunning{J{\o}rgensen, J.~K. et al.}\maketitle%\nolinenumbers

\section{Introduction} \label{sec:intro}
The chemistry that occurs during the formation of stars and planets plays a critical role in determining the eventual architectures of planetary systems, the composition of exoplanet atmospheres, and perhaps the inventory of organic molecules delivered to planets that may serve as prebiotic building blocks for life \citep[e.g.,][]{oberg21}. Characterizing the molecular complexity that can arise around young stars systematically and relating it to the physical evolution of the stars and the conditions in their ambient environment is therefore paramount. We present here an overview of a large program using the Atacama Large Millimeter/submillimeter Array (ALMA) to characterize the complex chemistry of a sample of 11 deeply embedded solar-type protostellar regions known to harbor warm gas rich in complex organics. 

In the earliest stages, during the epoch when a cold dense molecular cloud core collapses to form embedded protostars and protoplanetary disks, simple molecules such as CO, CO$_2$, H$_2$O, CH$_4$, and NH$_3$ are frozen out as ices on the surfaces of dust grains \citep[e.g.,][]{boogert15}. From these molecules, solid-state reactions can lead to the formation of more complex species, including complex organic molecules (COMs) typically defined as carbon-bearing species with six or more atoms \citep[e.g.,][]{herbst09}. During the cold stages, these species can be traced in the gas phase if they are injected into it through nonthermal desorption processes, as observed toward some prestellar cores \cite[e.g.,][]{bacmann12,vastel14,jimenezserra16}, or directly as ices through their infrared signatures recently made feasible through sensitive \textit{James Webb} Space Telescope (JWST) observations \citep[e.g.,][]{yang22,mcclure23}. However, only the most common complex organics can be reliably measured in this way because of their low column densities in the cold gas phase and the difficulties in disentangling the broad, overlapping, ice features at infrared wavelengths. Once the material falls in toward the spatial scales of the young protostar and its protoplanetary disk, the ices sublimate at the elevated temperatures of one to a few hundred kelvins resulting from heating by the young star itself. This injects them into the gas phase in the so-called ``hot corino'' where they can be readily observed by ALMA \citep[e.g.,][]{jorgensen20}. ALMA therefore provides a unique opportunity to map the full volatile composition of the material that is accreted onto protoplanetary disks.

From its earliest science verification days and throughout its first decade of operations, ALMA has demonstrated its powerful ability to characterize the complex chemistry that occurs around young solar-type stars. In addition to identifying several new hot corinos around such sources \citep[e.g.,][]{imai16,oya17,yang20}, it has provided several detections of complex organic species that were until then only detected toward selected high-mass star-forming regions (e.g., glycolaldehyde, CH$_2$(OH)CHO, \citealt{jorgensen12}; methyl isocyanate, CH$_3$NCO, \citealt{ligterink17,martindomenech17}), and even some that were seen for the first time in the interstellar medium (e.g., methyl chloride, CH$_3$Cl, \citealt{fayolle17}; glycolonitrile, HOCH$_2$CN, \citealt{zeng19}; and 3-hydroxypropenal, HOCHCHCHO, \citealt{coutens22,muller24}). ALMA has also enabled systematic characterization of the isotopic compositions of complex organic species \citep[e.g.,][]{jorgensen18}. 

ALMA's strength for these studies lies in several aspects: its high angular resolution allows us to zoom in on the hot ($T > 100$~K) gas surrounding the protostars where the ices are fully sublimated. This angular resolution reduces ambiguities about the origin of the emission and, particularly in regions with more extended and kinematically complex molecular emission, it enables studies of specific positions where narrower lines are identified, reducing spectral confusion (see, e.g., the example of the ``off position'' of the ``B component'' of IRAS~16293-2422 studied in the majority of papers from the Protostellar Interferometric Line Survey (PILS) program; \citealt{jorgensen16}).

The other key aspect of ALMA is its high sensitivity: a 30-minute observation of on-source integration time (corresponding to equal amounts of calibration and science time) delivers an RMS of 3~mJy~beam$^{-1}$ per 1~km~s$^{-1}$ channel at 300 GHz ($\approx0.16$~K for a 0.5$''$ beam). This sensitivity is sufficient to reach the line confusion limit (where the line density is so high that line
emission is detected in every single channel) for even some low-mass protostars, while still covering a wide bandwidth that enables systematic, unbiased inventories of rotational molecular transitions emitting at these frequencies, i.e., without prior assumptions about what molecules are present and under what physical conditions. Such unbiased surveys are critical for accurately determining excitation temperatures and column densities of these species, as well as to account for and predict the emission from fainter lines needed to identify rarer molecular species.

Although ALMA has provided important new insights into the complex chemistry of individual protostars, studies of the chemistry of embedded protostars have so far been limited to large spectral surveys of a few well-known sources or programs targeting specific sets of species. Consequently, more systematic overviews are still missing, leaving several important unanswered questions: in particular, it is not understood to what extent the chemistry of complex organic molecules is affected by environmental conditions (e.g., the temperatures, densities, and radiation fields) during the cold phases of the molecular clouds. One question, for example, is whether the chemistry of protostars formed in isolation reflects accretion over longer periods at low temperatures compared to protostars from clustered environments, as suggested by observations of water deuteration \citep{jensen19}. If so, this could affect the abundances of complex organics formed through grain-surface processes. It also remains unclear how protostellar physical evolution may affect the chemistry, i.e., the chemical processes and resulting molecular composition. There are further questions such as, for example, whether chemistry is affected by how material is accreted toward the disk scales and what role, if any, the properties and potential variability of the young stars play.

To address these questions, the Large Program called ``Complex Organic Molecules in Protostars from ALMA Spectral Surveys'' (COMPASS) carried out unbiased spectral surveys of 11 low-mass protostellar sources in diverse environments and at different evolutionary stages. In this paper, we present an overview of COMPASS, the observing strategy, and the targeted sample of sources, including brief overviews of their properties from the literature, as well as zeroth-order comparisons of their observed line emission in COMPASS. This paper is accompanied by dedicated papers that describe the data reduction strategy \citep{compassDR} and discuss the approach to fitting the broadband spectra of the survey, including the first results on methanol isotopologs observed toward the Class~0 protostar BHR71-IRS1 \citep{coutens26compass}. In addition, a series of ``first look'' papers describe different aspects of the first data delivered to the survey \citep[][]{jeong26compass,yun26compass,nazari26compass,belloche26compass} and set the stage for science that can be carried out with the complete dataset of COMPASS.

\section{Observations} \label{sec:observations}
\begin{figure*}
\resizebox{\hsize}{!}{\includegraphics[]{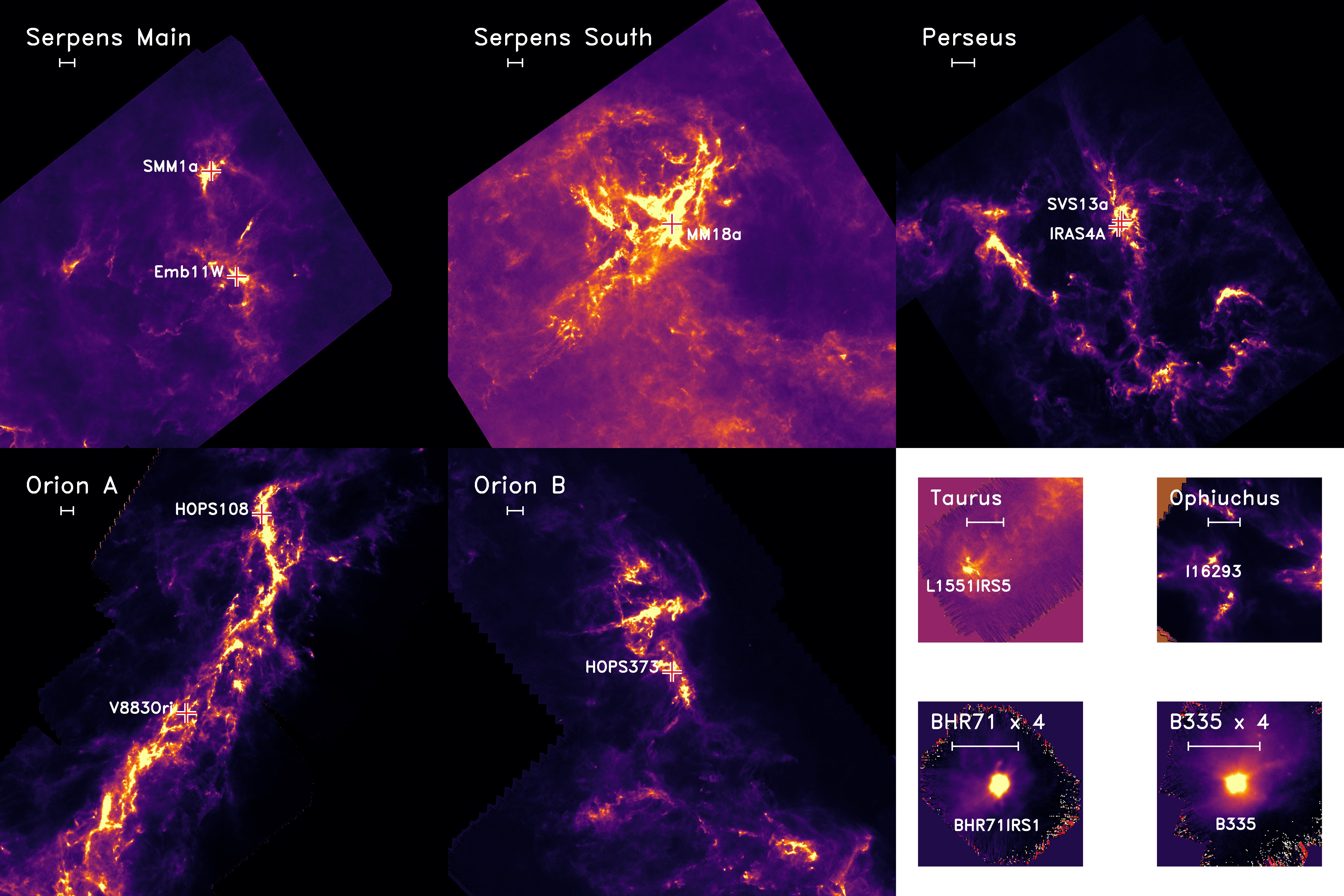}}
\caption{Column density maps where available from the \emph{Herschel} Gould Belt survey \citep{andre10}, supplemented by individual \emph{Herschel/PACS} 160~$\mu$m images for individual sources (Taurus: L1551-IRS5, BHR71, B335) from the \emph{Herschel} Science Archive. The scale bar in each plot indicates a linear scale of 0.8~pc, except for BHR71 and B335, where the images are zoomed in and the scale bar corresponds to 0.2~pc. Each source targeted here is indicated with the labels and for the sources in Orion, Serpens (South and Main), and Perseus also with a red and white plus sign.}\label{environments}
\end{figure*}

COMPASS was approved as a Large Program ($\sim$125~hours of ALMA 12~m array time) in 2022 in ALMA's Cycle 9 (proposal ID: 2022.1.00316.L\footnote{\url{https://erda.ku.dk/vgrid/COMPASS/}}). The program consists of an unbiased spectral survey of a sample of 11 protostellar sources from a diverse set of environments, ranging from highly clustered regions to individual, isolated, star-forming globules. Figure~\ref{environments} shows their distribution compared to their larger scale environments traced by the emission from cold dust as observed by \emph{Herschel}, including IRAS16293-2422 targeted by PILS: sources such as those in Serpens, the Serpens South region, Perseus, and Orion A are part of denser, clustered environments, whereas the others are located closer to the outskirts of larger-scale cloud environments (L1551-IRS5 in Taurus, HOPS 373 in Orion~B, and IRAS16293-2422 in Ophiuchus) or are part of isolated cores or globules (e.g., BHR71 and B335). The sources also represent different stages of protostellar evolution in traditional classification schemes based on their spectral energy distributions (see also Sect.~\ref{seds}). Table~\ref{table:sample} lists the sample of sources and Sect.~\ref{subsec:sample} discusses the literature for each of the sources and their key characteristics. This section presents the general considerations behind the observations and strategy and provides an overview of the approach to data reduction. We refer to \cite{compassDR} for detailed observation logs and full description of the approach to data reduction.
\begin{table*}
\caption{Targeted sample of sources.} \label{table:sample}
\centering
\begin{tabular}{llcccl}\hline\hline
Target & Abbr. name\tablefootmark{(a)}           & \multicolumn{2}{c}{Phase center} & HPBW\tablefootmark{(b)} & SIMBAD reference\tablefootmark{(c)}    \\
                             &   & RA (ICRS)    & DEC (ICRS)     & [min, max] ($''$) & \\ \hline
NGC~1333-SVS13a     & SVS13a        & 03:29:03.756 & $+$31:16:03.80 & [0.14,0.18] & \object{[SVS76] NGC 1333 13A1}   \\
NGC~1333-IRAS4A     & IRAS4A        & 03:29:10.432 & $+$31:13:32.12 & [0.26,0.48] & \object{[JCC87] IRAS 4A}         \\
L1551-IRS5          & L1551IRS5     & 04:31:34.160 & $+$18:08:04.72 & [0.13,0.16] & \object{L1551 IRS 5}               \\
HOPS 108            & HOPS108       & 05:35:27.084 & $-$05:10:00.07 & [0.28,0.41] & \object{HOPS 108}                  \\
V883-Ori            & V883Ori       & 05:38:18.100 & $-$07:02:26.00 & [0.28,0.41] & \object{V883 Ori}                  \\
HOPS 373-B          & HOPS373       & 05:46:30.905 & $-$00:02:35.20 & [0.28,0.42] & \object{NAME HOPS 373SW}           \\ 
BHR71-IRS1          & BHR71IRS1     & 12:01:36.499 & $-$65:08:49.38 & [0.36,0.44] & \object{[B2001b] IRS 1}          \\ 
Serpens emb 11 W    & Emb11W        & 18:29:06.720 & $+$00:30:34.30 & [0.26,0.44] & \object{[ECD2011] Ser-emb 11(W)} \\
Serpens SMM1a       & SMM1a         & 18:29:49.793 & $+$01:15:20.20 & [0.25,0.43] & \object{NAME Serpens SMM 1a}       \\ 
Serpens South-MM18a & MM18a         & 18:30:04.118 & $-$02:03:02.55 & [0.24,0.42] & \object{[MAM2011] SerpS-MM18}    \\
B335 IRS            & B335          & 19:37:00.894 & $+$07:34:09.59 & [0.23,0.49] & \object{NAME Barn 335 IRS}         \\ \hline
\end{tabular}
\tablefoot{
\tablefoottext{a}{Shortened name adopted in this paper.}
\tablefoottext{b}{Range of angular resolutions (Half Power Beam Width) for the imaging of the specific source. For more detailed logs see \cite{compassDR} and project website \url{https://erda.ku.dk/vgrid/COMPASS/specwindows.html}.}.
\tablefoottext{c}{Identifier in the SIMBAD database \url{https://simbad.cds.unistra.fr/simbad/}}.}
\end{table*}

\subsection{Targeted frequency range}
\begin{figure}
\resizebox{\hsize}{!}{\includegraphics{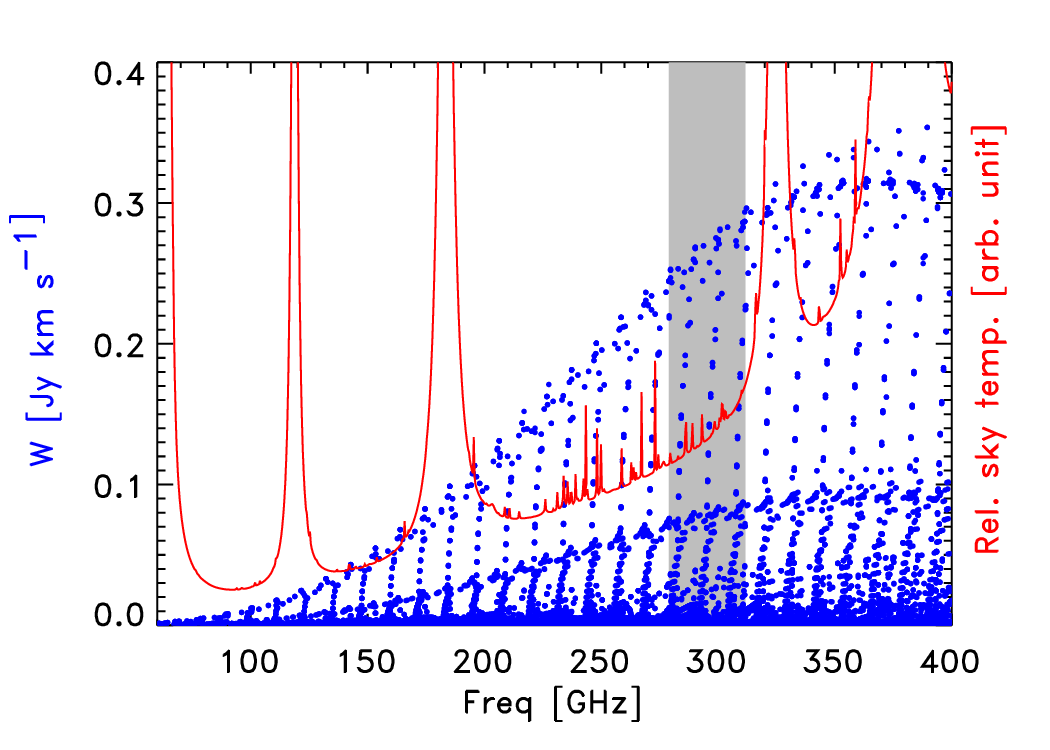}}
\caption{Comparison between estimated line strengths and achievable noise levels as a function of frequency. Blue symbols show the predicted methyl formate, CH$_3$OCHO, line strengths as a function of frequency assuming an excitation temperature of 150~K and a column density of $1\times 10^{16}$~cm$^{-2}$ uniformly extended over a 0.5$''$ beam. Only transitions with a predicted line opacity $\tau< 0.1$ are included in the plot and thus the predictions represent the variation of temperature-dependent, optically thin line emission as a function of frequency at this specific temperature. The red curve shows the variation in the sky temperature (arbitrary units) as a function of frequency for a PWV of 1.796~mm (corresponding to the fifth octile of weather conditions at Chajnantor). The sky temperature dominates the achievable noise level as a function of frequency for a given value of PWV and (ignoring variations in receiver temperatures and antenna efficiencies) scales with the achievable noise level for a given integration time. The targeted frequency range between 279.0 and 311.7~GHz (gray area) represents a sweet spot in terms of signal-to-noise ratio, as a compromise between the gain in line strengths and the loss in sensitivity due to the atmospheric conditions at increasing frequencies.}\label{frequencyselect}
\end{figure}

When deciding on the targeted frequency range to survey, we balanced several considerations. We produced synthetic spectra for typical temperatures of the hot gas seen toward solar-type protostars of $\sim 100-300$~K, adopting the column densities derived from the PILS program \citep{jorgensen16}. This provided an expected number of lines per bandwidth for the different species above a given observational noise level. Figure~\ref{frequencyselect} shows an example of such a calculation for one of the most common complex organics, methyl formate, CH$_3$OCHO; the plot shows the strength of optically thin ($\tau < 0.1$) lines as a function of frequency from 60 to 400~GHz (including ALMA's Bands 2--7) calculated for an excitation temperature of 150~K. As can be seen, the predicted line strengths increase steadily with a peak around 350~GHz.

This potential gain in signal needs to be balanced against the decrease in atmospheric transparency at higher frequencies. Figure~\ref{frequencyselect} also shows the variations in sky temperature over the same frequency range for a fixed amount of precipitable water vapor (PWV) of 1.796~mm corresponding to the fifth octile of weather conditions at Chajnantor (i.e., conditions reached more than 62.5\% of the available time) typically used for observations between 200 and 300~GHz. For such a PWV, the sky temperature increases by a factor of eight when going from 85 to 370~GHz (ALMA's Bands 3--7), with the most substantial rise above 300~GHz. The figure illustrates that there is a ``sweet spot'' between the gain in line strength as a function of frequency and the corresponding noise level in the frequency range of 279.0--311.7~GHz targeted in COMPASS. Obviously, observations at higher frequencies would likely be pursued in better weather conditions than the fifth octile (the upper part of Band 7 in the frequency range of 329--363~GHz targeted in PILS is typically in the fourth or third octile). However, the number of observable hours there is more limited.

The 33~GHz width of the targeted frequency range, and thus the number of individual observations required per source, was determined by considering the number of potentially detected species and the number of transitions per species. Investigating different setups, we found that a width matching that of the PILS survey was necessary to \textit{(i)} secure the presence of lines of all key complex organics, \textit{(ii)} ensure a sufficient number of lines for accurate determinations of excitation temperatures and column densities for the most common complex organics and their isotopologs \citep[e.g., based on and in comparison to][]{jorgensen18}, and \textit{(iii)} provide sufficient bandwidth to ensure that enough unblended lines are present for more common species and to search for rarer ones.

\subsection{Spatial and spectral resolution sensitivity}
For nine of the 11 targeted sources, we aimed for an angular resolution of 0.3--0.5$''$, corresponding to a typical linear size of 100--200~au at the distances of the targeted sources. This linear resolution is comparable to the estimated sizes of the regions of protostellar envelopes where the temperatures exceed 100~K and icy mantles are expected to sublimate (see also Sect.~\ref{seds}). Thus, the observations are expected to capture the full emission from molecules in this gas. Although they do not typically resolve the structures spatially, they provide sufficient resolution to separate this emission from more extended emission, for example, related to the outflows on larger scales. For ALMA, this range of angular resolution is typically obtained in configuration C-4 (baselines of 15--740~m), with some higher frequency observations obtained in configuration C-3 (baselines of 15--500~m). 

The exceptions to this choice are two sources, NGC1333-SVS13a and L1551-IRS5, which are known to be close binaries with both components showing emission from complex organics \citep[e.g.,][]{Bianchi2020_MNRAS498L87B,bianchi22svs13,Takakuwa2020_ApJ89810T}. These two sources were observed with an angular resolution of $\approx$0.15$''$, which is sufficient to resolve the binary components separated by 0.3--0.4$''$ and to identify the contributions to the integrated COM emission of each component. This angular resolution was achieved with the antennas in configuration C-6 (baselines of 15--2500~m).

In terms of spectral resolution, we aimed for a uniform frequency resolution of 488.281~kHz per channel (3840 channels per spectral window of 937.5~MHz in width). This results in a velocity resolution of 0.52--0.47~km~s$^{-1}$ per channel in the targeted spectral range. Although this is about a factor of two coarser than adopted for the PILS program, it was deemed sufficient for the targeted sources, all of which show line widths $\gtrsim$~2~km~s$^{-1}$, slightly wider than in the warm gas close to the protostellar B component of IRAS 16293-2422 ($\approx$1~km~s$^{-1}$) studied in several papers in connection with the PILS program.

The targeted sensitivity was set at 2.5~mJy per 1~km~s$^{-1}$ channel, driven by the desire to detect species with abundances down to 0.001--0.01\% relative to methanol for the targeted sources based on synthetic spectra. For comparison, many oxygen-bearing COMs with two carbon atoms (ethanol, methyl formate, acetaldehyde) were found in PILS to have abundances at a level of $\sim$~1\% relative to methanol \citep{jorgensen18}. With typical isotope ratios for D/H and $^{13}$C/$^{12}$C at the level of 1 to a few percent, this sensitivity is sufficient to detect the rare isotopologs of most of these species or otherwise provide meaningful upper limits. The actual sensitivity achieved matches or even exceeds these goals for all observed spectral settings \citep[see][and the COMPASS website for full observing logs]{compassDR}.

\subsection{Overview of observing executions}
The observations were carried out between January 8, 2023 and August 13, 2024. With the setup described above, the observations were split into 63 individual scheduling blocks (SBs), each targeting one source in one of nine different tunings except for the three sources in Orion and the three sources in Serpens, which were observed in two sets of clustered sources in common SBs to minimize overheads. To achieve the requested RMS, each SB was executed one to three times (one execution is called an ``execution block,"or ``EB''). Each EB was quality assessed by the observatory (QA0 and QA2) before being delivered. In total, 107 EBs were executed and passed through the QA0 to QA2 process as the basis for the program\footnote{For reference, for each EB the calibrated measurement set (the minimum product usable for imaging to produce products for science use) had an average size of $\approx$125 GB.}. After delivery, one scheduling block was found to have a faulty tuning related to a Phase~2 setup issue in the ALMA-OT software; after QA3, it was reentered into the observing queue and repeated. 

\subsection{Data reduction}
The data reduction procedure is described in detail in \cite{compassDR} and here just the main points are summarized. We start from calibrated measurement sets based on the ALMA pipeline and restored and delivered by the European ARC at ESO. For further processing, we use the IMAGER package based on the GILDAS software\footnote{IMAGER: \url{https://imager.oasu.u-bordeaux.fr}\\ GILDAS: \url{https://www.iram.fr/IRAMFR/GILDAS}}. The COMPASS pipeline IMAGER scripts are available through Zenodo \citep{compassdrscripts}. As discussed in \cite{compassDR}, no discernible differences are found between the results of the reduction in IMAGER and by using CASA \citep{casa2022}. For our data, IMAGER offers significantly higher speed: it performs operations using internal buffers, avoiding as much as possible saving intermediate data on disk, which otherwise is a significant bottleneck. From the calibrated measurement sets, we perform self-calibration on the line and continuum tables. This typically improves the continuum S/N, but does not strongly affect the line data. An initial estimate of the continuum flux is obtained using IMAGER's native algorithm for identifying line-free channels. The continuum and line cubes were Fourier transformed and deconvolved using the Hogbom clean algorithm. The continuum images were found to be limited by the dynamical range; for these, the stopping criterion for the deconvolution is that the total flux in the image converges to a constant value. The line data cubes were found to be thermal noise limited, and the stopping criterion for the deconvolution is that the final noise is the expected thermal noise. After inspecting the line cubes, we found that the initial automated continuum identification led to oversubtraction of the continuum toward the line-rich regions at the level of the line RMS noise. We therefore reimaged the cubes without continuum subtraction and thereafter applied a statistical pixel-by-pixel subtraction of the continuum level following the approach described in \cite{jorgensen16}.

\section{Sample}\label{subsec:sample}
The 11 targeted sources were selected from nearby ($d < 500$~pc), low-mass protostellar sources known to show compact emission from complex organic molecules from previous studies in the literature (see Sects.~\ref{firstsource}--\ref{lastsource} for details). We targeted sources with sufficiently bright lines so that species with abundances down to at least 0.01\% relative to methanol would be detectable, according to synthetic spectra based on the PILS data for the B component of IRAS~16293-2422 (IRAS16293B) scaled to individual sources. A further down-selection was made by excluding sources with broad line profiles (above 5~km~s$^{-1}$, e.g., L483; \citealt{oya17,jacobsen19}) that would complicate the identification of fainter species due to blending. We also limited the number of northern sources (e.g., excluding sources such as NGC1333-IRAS2A) due to scheduling considerations. This section provides a brief discussion of the literature for each of the targeted sources, organized by cloud environment. We describe the key physical characteristics and chemical signatures of each source, in particular focusing on the identification and presence of complex organics. For each region, we also summarize the most recent estimates of their distances and the values adopted here (see Table~\ref{SEDresults}).

\subsection{Descriptions of the targeted sources}
\subsubsection{Perseus: NGC1333-IRAS4A and -SVS13A}\label{firstsource} 
\parhead{NGC1333-SVS13A} The NGC1333 region is notable for its high concentration of young stars and Herbig Haro (HH) objects, among which HH~7--11 \citep{herbig74} are some of the most extensively studied. HH~7--11 have long been associated with the infrared source SVS13 \citep{strom76ngc1333} or with nearby embedded sources \citep[e.g.,][]{reipurth93,chini97}. (Sub)millimeter single-dish and interferometric observations have revealed the presence of several protostars in the immediate vicinity, including the most luminous source traditionally referred to as SVS13, but more recently as SVS13a or SVS13A\footnote{Sometimes also referred to as V512 Per and Per-emb~44.}, and a more deeply embedded source (invisible at shorter wavelengths) SVS13b at a distance of 15$''$ (4350~au). A radio source, VLA4, was found to be associated with SVS13a \citep{rodriguez97} and later resolved into a 0.3$''$ (90~au) binary source (VLA4A and VLA4B) at cm wavelengths \citep[e.g.,][]{anglada00,tobin16binary}.

The detection of high-excitation millimeter lines of water (HDO) in IRAM single-dish observations \citep{codella16svs13} and direct imaging of compact emission from glycolaldehyde (CH$_2$(OH)CHO) by the CALYPSO program \citep{desimone17} provided the first indications of a hot corino toward SVS13a. Subsequently, the two components of SVS13a were separately imaged at shorter (submillimeter) wavelengths and were found to be associated with the emission of complex organic species \citep{diazrodriguez22,bianchi22svs13}. Systematic surveys from multiple single-dish and interferometic programs have  built up a detailed molecular inventory of SVS13a, including studies of sulfur chemistry, abundances of complex organics, and deuteration \citep[e.g.,][]{codella21,belloche20,bianchi22ch3cn}. Recent work has also focused on understanding the physical origin of emission from these species, including the importance of accretion shocks triggered by streamers of accreting material \citep[e.g.,][]{bianchi23,hsieh23,hsieh24}.

\parhead{NGC~1333-IRAS4A}
NGC~1333-IRAS4A (IRAS4A) was first identified as a candidate protostellar source from IRAS observations of the NGC~1333 cloud \citep{jennings87} and was termed NGC~1333-IRAS4. Due to the presence of two separate water masers, \citeauthor{jennings87} speculated that the infrared IRAS source reflected two separate sources, heavily embedded, which was confirmed by submillimeter imaging \citep{sandell91}. Subsequent high-angular resolution imaging at millimeter wavelengths \citep{looney00} revealed that both of these sources, IRAS4A and IRAS4B, are multiple in nature. IRAS4A is resolved into two separate components at a separation of 1.8$''$ or $\approx$~500~au at the most recent distances of NGC1333 from Gaia.

\cite{blake95} performed the first systematic molecular line survey of IRAS4A (and IRAS4B) utilizing the James Clerk Maxwell and Caltech Submillimeter Observatory single-dish telescopes between 200 and 400~GHz. One of the key results from that study was the relatively low abundance of several molecular species, possibly indicating the freeze-out of molecules as ices onto dust grains in the large-scale cold envelope. While the protostellar outflows of IRAS4A clearly play an important role in regulating the molecular emission on the single-dish scales of IRAS4A \citep{blake95,maret04,hotcoresample,prosacpaper}, its luminosity and the predicted amount of warm envelope material ($>100$~K) resulting \citep{maret02} made it an excellent candidate for searches for complex organic molecules in its protostellar hot corino. Indeed, \cite{bottinelli04n1333i4a} reported the detection of several lines of methyl formate, formic acid, and methyl cyanide (CH$_3$OCHO, HCOOH, and CH$_3$CN) from single-dish observations. Millimeter wavelength interferometric observations revealed that the bulk of this COM emission \citep[e.g.,][]{taquet15,lopezsepulcre17,sahu19}, as well as compact, thermal, H$_2$O emission \citep[e.g.,][]{persson12}, is associated with the northwestern component of IRAS4A, referred to as IRAS4A2, while the southeastern component, IRAS4A1, dominates the dust continuum emission at these wavelengths. However, centimeter wavelength interferometric observations \citep[e.g.,][]{desimone20myth,desimone22dust} show methanol emission toward both components, indicating that the absence of COM emission toward IRAS4A1 at higher frequencies may result from a high dust opacity.

\parhead{Distance} There are 18 sight lines in \cite{zucker20} within 5$^\circ$ of IRAS4A and SVS13a with an average distance of 282$\pm 29$~pc. Within 1$^\circ$, four of those remain with distances of 276--291~pc and typical uncertainties of $\pm$10~pc. Based on Gaia DR2 measurements, \cite{zucker18} estimates a distance for NGC1333 of 299~pc, while \cite{ortizleon18} finds a distance of 293$\pm22$~pc for NGC1333. We adopt a distance of 290~pc.

\subsubsection{Taurus: L1551-IRS5} 
\parhead{L1551-IRS5} 
L1551-IRS5 (L1551IRS5) is an embedded protostellar system in the Taurus star-forming region. The system is composed of two small dust disks (radius $\lesssim$10 au at 1.3\,mm) separated by $\approx$50\,au \citep{rodriguez98,Lim2016_ApJ826153L,CruzSaenzdeMiera2019_ApJ882L4C}, a circumbinary disk or flattened inner envelope with a radius of $\sim$150\,au \citep{CruzSaenzdeMiera2019_ApJ882L4C,Takakuwa2020_ApJ89810T}, and a $\sim$1--2~$M_\odot$ envelope on larger scales \citep{jorgensen02,Chou2014_ApJ79670C}. It drives a bipolar outflow \citep{Snell1980_ApJ239L17S,Wu2009_ApJ698184W} and two jets detected at optical/near-infrared wavelengths \citep{Mundt1983_ApJ274L83M,Liseau2005_ApJ619959L} and at radio frequencies \citep{Rodriguez2003_ApJ586L137R,Lim2016_ApJ826153L}.

Throughout the past four decades, molecular emission around L1551IRS5 has been observed using multiple single-dish and interferometric facilities, covering multiple transitions from several simple molecules \citep[e.g., CS, CO, HCO$^+$, and some of their isotopologues;][]{Fridlund2002_AA382573F}, as well as CH$_3$OH \citep{White2006_ApJ651L41W} on large scales. ALMA observations by \cite{Bianchi2020_MNRAS498L87B} provided the first detections of rarer species, including two methanol isotopologs ($^{13}$CH$_3$OH and CH$_2$DOH), methyl formate (CH$_3$OCHO), and ethanol (CH$_3$CH$_2$OH), while NOEMA observations recovered emission from HDO and H$_2${$^{18}$}O \citep{Andreu2023_AA677L17A}. The ALMA observations by \cite{cruzsaenzdemiera25} with an angular resolution of $\sim 0.2''$ separated the emission around the two protostars, resolving emission from several additional COMs. Recently, a more comprehensive chemical survey using the IRAM~30-m telescope in the 2 mm and 3 mm bands \citep{Marchand2024_AA687A195M} revealed emission from 75 different species (37 main and 38 secondary isotopologues), including 10 COMs.

\parhead{Distance} The closest sight line from \cite{zucker20} toward L1551-IRS5 is 6$^\circ$ away, and eight are within 10$^\circ$. The mean distance for those is 145~pc with a standard deviation of 12~pc (median 149~pc). We adopt 145~pc as the distance for L1551-IRS5.

\subsubsection{Orion A: HOPS~108 and V883~Ori} 
The integral-shaped filament (ISF) within the Orion A molecular cloud comprises the Orion Molecular Clouds OMC-1, OMC-2, and OMC-3. \citet{mezger90} identified multiple regions in OMC-1 and OMC-2 at submillimeter/FIR wavelengths, several of which are considered protoclusters \citep{shimajiri08,lopez-sepulcre13,kainulainen17}. 

\parhead{HOPS 108}
Among the protoclusters in Orion A, OMC-2/FIR4 has drawn particular interest as a potential analogue of the Sun's birth environment due to its clustered nature and high ionization level \citep{ceccarelli14}. \emph{Herschel} observations identify HOPS~108 as the most luminous protostar associated with OMC-2/FIR4 \citep{furlan14}. \citet{tobin19} studied the region at high angular resolution (0.1$''$) using ALMA and the VLA, identifying several protostars including HOPS~108. They found that HOPS~108 is likely a protostar of low- to intermediate-mass in its main accretion phase. 

The chemistry of OMC-2/FIR4 has been studied on arcsecond scales within the NOEMA SOLIS program \citep{ceccarelli17}, confirming high levels of irradiation \citep{fontani17,favre18} and detecting a possible SiO jet \citep{lattanzi23}. ALMA observations by \citet{sato23} highlight the complexity of the region and provide evidence for the interaction between HOPS~108 and the energetic outflow from protostar HOPS~370 (in OMC-2/FIR3), confirming earlier suggestions \citep[e.g.,][]{osorio17}. Compact warm methanol emission toward HOPS~108 was detected by \citet{tobin19}, and the presence of a hot corino was confirmed with the detection of many other COMs in ALMA observations by \cite{chahine22}.  

\parhead{V883 Ori}
V883 Ori (also known as HOPS 376) is an FU Orionis object in the Orion A molecular cloud L1641, located south of the Orion Nebula Cluster. The onset of its outburst was not directly observed, but likely occurred before 1888, based on the associated reflection nebula \citep{pickering1890} and its near-IR spectrum \citep{connelley18}. The bolometric luminosity is approximately 200 $L_\odot$ and may have been higher in the past \citep{strom93,sandell01fuori,furlan16}. V883 Ori is a 1.3 $M_\odot$ young stellar object in transition between the Class I and Class II stages, with a well-developed Keplerian-rotating disk \citep{cieza16} with a dust mass of $\sim$0.6 $M_\odot$ \citep{cieza18}, but the envelope is still present and an outflow has been observed in CO emission \citep{ruizrodriguez17,white19,ruizrodriguez22}.  

The water snowline was initially inferred at a radius of around 40~au based on a change in the continuum spectral index \citep{cieza16}, but recent water observations (HDO and H$_2^{18}$O) suggest a midplane snowline radius of 80~au \citep{tobin23}. The first detection of a COM (CH$_3$OH) was presented in \citet{vanthoff18_v883ori}, followed by detections of additional COMs such as CH$_3$OCHO, CH$_3$CHO, and CH$_3$CN in ALMA Band 7 observations \citep{lee19-v883ori}. COM emission extends to distances of $\sim$120 au from the central star and displays a velocity structure consistent with the Keplerian rotation traced by C$^{18}$O \citep{cieza16}, supporting a disk origin. Subsequent chemical surveys in ALMA Bands 3 and 6 \citep{lee24,yamato24,jeong25v883ori,fadul25} have revealed that the disk is chemically rich in oxygen-bearing COMs, whereas nitrogen-bearing species other than CH$_3$CN appear to be deficient. In terms of deuteration, the source interestingly shows a D/H ratio that is on the low side for methanol \citep{zeng25} but on the high side for water \citep{tobin23} compared to other protostellar sources. 

\parhead{Distance} Twelve Gaia sightlines of \cite{zucker20} within 3$^\circ$ of HOPS 108 suggest a distance of 417$\pm 23$~pc and nine sightlines toward V883Ori indicate a distance of 438~pc, but with a large uncertainty of $\pm 45$~pc. For V883Ori, \cite{lee19-v883ori} analyzed the Gaia distances toward 47 members of the L1641 cluster and found a distance of 388$\pm 16$~pc. \cite{grossscheld18} analyzed the 3D structure of Orion A and reported a distance of 393$\pm$25~pc toward the ISF, but a slightly larger distance of 424$\pm$42~pc toward L1641 around V883Ori. While the uncertainties mean that the distance estimates are all in mutual agreement, it is perhaps good to note that there could be systematic errors that are larger than for other regions. However, as the distances for many of our purposes are less critical, we adopt distances of 388~pc for V883Ori and 393~pc for HOPS~108 for consistency with \cite{lee19} and \cite{chahine22}, respectively.

\subsubsection{Orion B: HOPS~373} 
\parhead{HOPS~373} 
HOPS 373 is a protostar located in the NGC~2068 region within Orion B that was initially detected through its large ($>$1\arcmin) outflow visible in CO and HCO$^+$ emission \citep{gibb00_hops373}. A continuum source was subsequently identified in submillimeter observations taken with SCUBA on the JCMT \citep{johnstone01,motte01orion,phillips01}. HOPS 373 is classified as a so-called ``PACS Bright Red Source'' based on its steeply rising SED from 24~$\mu$m to 70~$\mu$m \citep{stutz13}, a feature of some of  the youngest protostars.
Monitoring in the infrared (NEOWISE) and submillimeter (JCMT) shows that the source is variable \citep{johnstone18,contreras-pena20,lee21,park21}; recently, it showed a $\sim$30\% increase in 850 $\mu$m brightness \citep{yoon22}. 

Interferometric millimeter observations resolved the source into two components, HOPS 373 NE and HOPS 373 SW, separated by $\sim$1500 au \citep{tobin15orion,tobin20}. ALMA observations revealed a hot corino toward HOPS 373 SW with high abundances of single, doubly, and triply deuterated methanol \citep{lee23}. In contrast, only C$^{17}$O was detected toward the NE source, which implies a cold and quiescent environment. 

\parhead{Distance} \cite{kounkel17} gives a distance of 388$\pm$10~pc toward NGC~2068 based on VLBA measurements. The average distance from ten sight lines within 3$^\circ$ of HOPS~373 in \cite{zucker20} is 428$\pm 15$~pc, consistent with the estimate of 417$\pm$5~pc in \cite{kounkel18}. \cite{kounkel18} attribute the nearer VLBA distance to small-sample bias and binarity effects. We therefore adopt the distance of 417~pc here.

\subsubsection{Serpens Main: Serpens emb 11 W and SMM1a} 
The Serpens molecular cloud (sometimes called Serpens Main to distinguish it from Serpens South, e.g., \citealt{lee14}) is an active star-forming region that hosts several deeply embedded protostars as imaged at (sub)millimeter wavelengths \citep[e.g.,][]{casali93,testi98,testi00,hogerheijde99}. A high level of outflow activity is observed, reflecting the youth of the region \citep[e.g.,][]{kristensen10serpens,goicoechea12,dionatos13,hull16,tychoniec19}. Line emission on larger scales of simple diatomic and triatomic species has also been studied to understand the global properties of the cloud and implications for star formation within it, including its filamentary structure \citep[e.g.,][]{graves10,lee14}.

\parhead{Serpens SMM1a} 
Of the embedded protostars in the Serpens Main region, SMM1 is the brightest source \citep[e.g.,][]{harvey84,casali93,enoch07}, and is classified as a Class~0 protostar based on its spectral energy distribution \citep[e.g.,][]{casali93,hurt96,hogerheijde99}. High-resolution continuum observations by \citet{choi09} reveal that SMM1 consists of multiple sources within a few thousand au, with Serpens SMM1a dominating in luminosity \citep{hull17,tychoniec19}. SMM1a was first identified as rich in complex organic molecules by \cite{oberg11serpens}. Later studies \citep{ligterink21,ligterink22} demonstrated the presence of nitrogen-containing prebiotic species, including methyl isocyanate (CH$_3$NCO) and glycolonitrile (HOCH$_2$CN).

\parhead{Serpens emb 11 W} 
Serpens emb 11 (Emb11) was identified as a Class I protostar in Cluster B of the Serpens Molecular Cloud using Bolocam 1.1 mm and \textit{Spitzer} observations \citep{enoch09}. Given its bolometric temperature of 77~K, close to the Class 0/I boundary at 70~K \citep{andreppiv}, Emb11 can be considered as a source in transition from Class 0 to Class I. CARMA observations resolved Emb11 into a binary with a separation of 2.1" ($\sim$900 au) between the eastern and western components, Emb11E and Emb11W \citep[][]{enoch11}. Near-IR images show outflow emission, but could not constrain the driving sources \citep{djupvik16}. More recent ALMA observations of $^{12}$CO suggest that both components drive  molecular outflows \citep{martin-domenech21}. Although emission from small molecules such as C$^{18}$O, CS, and H$_2$CO is observed toward both components, COMs are detected only toward Emb11W \citep{martin-domenech21,vanthoff24}. The COM emission is unresolved in the $\sim$0.5" ALMA beam. Signs of rotation are observed for C$^{18}$O, but current observations cannot determine whether a rotationally supported disk is present \citep{martin-domenech21}. 

\parhead{Distance} Two sight lines in \cite{zucker20} lie within 1.5$^\circ$ of Serpens emb 11 W and Serpens SMM1a, with distances of $425^{+12}_{-16}$ and $490^{+14}_{-11}$~pc. The former is 0.4$^\circ$ from both sources; the latter is 1.3$^\circ$ from Serp SMM1a and 0.5$^\circ$ from Serp emb 11W. Both are associated with Serpens Main for which \cite{herczeg19} derived a distance of 445$^{+30}_{-60}$~pc. \cite{ortizleon18} used VLBA parallaxes of radio continuum sources and derived a distance of 436.0$\pm$9.2~pc for Serpens Main. We adopt the latter distance for these two sources.

\subsubsection{Serpens South: MM18a}
\parhead{Serpens South MM18a} 
SerpS-MM18 refers to the millimeter continuum source identified from single-dish observations of the Serpens South cluster of the Aquila Rift complex \citep{maury11,Nakamura2011a}. It consists of multiple sources, including SerpS-MM18a and -MM18b, with the latter located $10''$ to the southwest of SerpS-MM18a \citep{Maury2019}. SerpS-MM18a, also referred to as CARMA~7, stands out as the brightest millimeter continuum source of the region \citep{Plunkett2015a} and has been proposed to be associated with a Class 0 protostar  \citep{gutermuth08,maury11}. The observations of COMPASS are centered on MM18a, placing MM18b at the edge of the primary beam of the ALMA observations.

MM18a is found to be surrounded by a massive dust disk and envelope \citep{Plunkett2018} and to drive a highly collimated bipolar outflow that extends over $\sim$ 0.16 pc \citep{Plunkett2015b}. Several knots are seen along the outflow, suggesting episodic events attributed to variations in the accretion rate of mass to the protostar \citep{Plunkett2015b}. \cite{Podio2021} made the first detection of collimated high-velocity jets in SiO and SO. They observed that the jet exhibited a slight precession, a phenomenon also seen in CO emission \citep{Plunkett2015b}. In terms of its chemistry, MM18a is somewhat less well studied, but based on data from the CALYPSO survey, \cite{belloche20} found that the source is a chemically rich hot corino-like object.

\parhead{Distance} There are ten sight lines from \cite{zucker20} within 4$^\circ$ of Serpens South MM18, five of which are within 1$^\circ$. The distances range from 439 to 556~pc with a median of 489~pc and a mean of 494~pc (standard deviation of 42~pc). \cite{anderson22} derived a distance of 455$\pm$50~pc from Gaia Early Data Release 3 for Serpens South. \cite{ortizleon23} derived a distance based on parallax measurements of 22~GHz masers associated with the protostar CARMA-6 of 440.7$\pm 4.6$~pc. We adopt 441~pc as the distance of Serpens South MM18. 

\subsubsection{Isolated cores: B335 and BHR71-IRS1}\label{lastsource} 
\parhead{B335} B335 is one of the prototypical objects for demonstrating the formation of stars from isolated cores. The detection of a cold and compact source in the far-infrared toward the B335 Bok globule by \cite{keene83} provided some of the first evidence for the formation of solar-type stars from such isolated dense dust clouds. Later studies of molecular line emission toward B335 led to the first detection of kinematical signatures indicative of infalling motions, with yet more recent studies comparing the kinematical structures for models of infall \citep[e.g.,][]{zhou93,choi95,evans05b335,evans15}. With its relatively simple geometry, B335 was also an early target for searches for embedded protostellar disks in (sub)millimeter continuum observations \citep[e.g.,][]{harvey03b,prosacpaper,evans23}. Although such comparisons suggest the presence of compact dust emission that cannot be accounted for by models explaining the larger-scale infalling envelope, high-resolution kinematical studies do not find evidence for Keplerian rotation down to scales of $\sim 10$~au \citep{yen15b335,bjerkeli19,imai19,okoda22}. B335 has recently been found to vary in flux at infrared wavelengths \citep{kim24}, extending previous evidence for variability in the source from, for example, the outflow morphology on both large \citep[e.g.,][]{reipurth92b335} and small scales \citep{reipurth02,bjerkeli19,kim24}.

Emission from complex organic molecules and water vapor is seen in a compact region around the central protostar \citep{imai16,imai19,okoda22,jensen19,jensen21observations}. Similarly to other isolated protostellar cores, the relative abundances of deuterated water isotopologs (HDO, D$_2$O) compared to normal water suggest  inheritance from cold molecular clouds that has evolved through prestellar stages on longer timescales than protostars formed in more clustered environments \citep[e.g.,][]{jensen19,jensen21observations}. The recent increase in luminosity appears to have caused complex organic molecules to sublimate into the gas phase where they have not yet had time to re-freeze onto dust grains \citep{lee25}.

\parhead{BHR71-IRS1} BHR71-IRS1/2 is a deeply embedded protostellar system in a southern Bok globule first cataloged by \cite{bourke95}. Early CO observations from the SEST telescope \citep{bourke97} revealed a highly collimated outflow emerging from the center of the core consistent with the presence of a Class 0 protostar. Deep K$'$-band imaging combined with ISO mid-infrared observations \citep{bourke01} suggested the presence of a binary system with a separation of about 16$''$ (2800--3200~au for typical distances; see below). Each component drives separate but overlapping outflows, as observed with the APEX telescope \citep{parise06bhr71}. These and other early single-dish observations \citep[e.g.,][]{garay98} show strong localized emission of CH$_3$OH and SiO toward the outflow, indicating active shock chemistry \citep[e.g.,][]{schilke97}.

With its location in the far southern hemisphere, BHR71 has been a prime ALMA target, particularly in studies of the kinematic signatures of the inner envelope and putative circumstellar disks \citep[e.g.,][]{tobin19bhr71,yang20,gavino24}. The system shows a complex velocity structure, with gradients both on large scales and in the immediate vicinity of each component. While small-scale ($<1000$~au) velocity gradients are perpendicular to protostellar outflows, they show no clear evidence of Keplerian rotation at the smallest resolved scales, leaving the disk sizes unconstrained. Emission from complex organic molecules beyond CH$_3$OH was first identified toward BHR71-IRS1 by \cite{yang20}.

\parhead{Distance} The distances to B335 and BHR71IRS1 were recently reviewed by \cite{evans23} and \cite{gavino24}, respectively. In brief, B335 shows a brightened rim caused by scattering light from a nearby A2 star, HD184982 \citep{watson20}. GAIA DR3 measurements demonstrate that that particular star is located at a distance of 164.5~pc, which \cite{evans23} consequently adopt as the distance of the core. BHR71IRS1 is associated with the Southern Coalsack for which distances in the range 150--200~pc have been quoted. \cite{voirin18} derive an average distance of 176~pc that \cite{gavino24} adopt. We follow \cite{evans23} and \cite{gavino24} and adopt distances of 164.5~pc and 176~pc for B335 and BHR71-IRS1, respectively.

\subsection{Spectral energy distributions}\label{seds}
\begin{figure*}
\resizebox{\hsize}{!}{\includegraphics[]{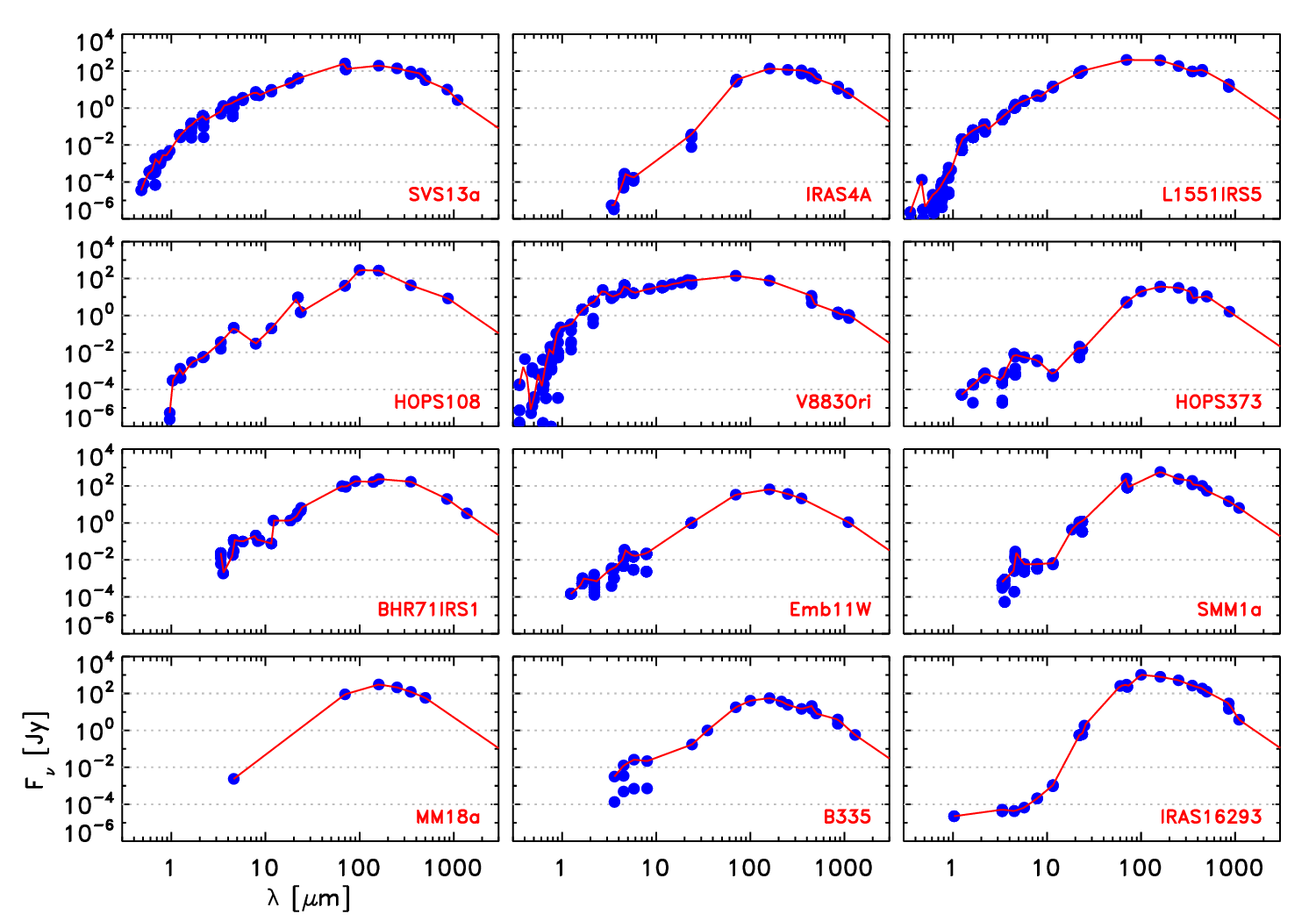}}
\caption{Spectral energy distributions for the 11 sources targeted by COMPASS and compared to IRAS16293-2422 (bottom right panel). The red line in each panel represents a spline fit to the measured photometric data points.}\label{sedfig}
\end{figure*}

As part of a review of the background literature for the targeted sources, we collected the most recent information on the spectral energy distributions (SEDs) for each source. As a starting point for the SED compilations, we utilized the database provided by the NASA/IPAC Infrared Science Archive (IRSA\footnote{\url{https://irsa.ipac.caltech.edu}}) that queries the main all-sky catalogs from visible through infrared wavelengths to (sub)millimeter wavelengths. Each SED was inspected carefully: there were some cases where several different infrared catalog entries were attributed to one source within a given search radius, and for those, we evaluated which entries corresponded to the actual source. Conversely, we did not filter out measurements for the same sources from different catalogs representing the same surveys (e.g., WISE or \textit{Spitzer}), but rather allowed them to enter the SEDs and weighted them according to the measurement uncertainties reported in the respective catalogs. Some entries were excluded, for example, submillimeter measurements with very large beams encompassing extended cloud regions (such as those from the \emph{Planck} mission). We also investigated whether photometry from single-dish measurements referred to peak flux densities (Jy~beam$^{-1}$) or integrated fluxes (in Jy), adopting integrated fluxes when available. Reported measurements from interferometric observations were excluded, as they do not reflect the overall energy output of the protostellar sources. In addition to the fluxes from the IRSA queries, we included measurements from \emph{Herschel} surveys, in particular the \emph{Herschel} Gould Belt Survey \citep{andre10}, where available. An overview of the compiled SEDs is shown in Fig.~\ref{sedfig}.

Based on the recompiled SEDs, we recalculated the bolometric temperatures (defined as the temperature of a black body that emits with the same mean frequency as the target source; \citealt{myers93}) and bolometric luminosities. These calculations were carried out by first performing spline fits to the SED measurements and then numerically integrating across the wavelength range. For bolometric luminosities, we adopted the updated distances from the previous sections. The results are listed in Table~\ref{SEDresults}. As can be seen, the sample contains sources with bolometric temperatures ranging from around 30 to several hundred kelvins, that is, ranging from very deeply embedded Class~0 protostellar sources to quite evolved Class I protostars \citep[see, e.g.,][for a review of the classification systems for YSOs]{evans09}. In terms of luminosities, the sample includes sources ranging from one to several hundred $L_\odot$. These luminosities refer to the total luminosities of the systems, not to individual components within binary or multiple systems, for example.
\begin{table*}\centering
\caption{Source properties from SED fits.\label{SEDresults}}\centering
\begin{tabular}{lllllll}\hline\hline
Target & $d$\tablefootmark{(a)} & $T_{\rm bol}$\tablefootmark{(b)} & $L_{\rm bol}$\tablefootmark{(b)} & $r_{\rm 100 K}$\tablefootmark{(c)} & $N_{T > 100 K}$\tablefootmark{(c)} & Rel. strength\tablefootmark{(d)} \\ 
       & [pc] & [K] & $[L_{\odot}]$ & [au] & $[10^{23}~{\rm cm}^{-2}]$ & \\ \hline
SVS13a        & 290 & 199 & 47  & 99  & 3.1  & 1.8 \\
IRAS4A        & 290 & 27  & 9.1 & 60  & 10.3 & 2.1   \\
L1551IRS5     & 145 & 164 & 17  & 58  & 3.0  & 0.58 \\
HOPS108       & 393 & 53  & 46  & 136 & 10.7 & 11.5 \\
V883Ori       & 388 & 664 & 189 & 169 & 0.50 & 0.83 \\
HOPS373       & 417 & 36  & 4.9 & 43  & 9.0  & 0.96 \\
BHR71IRS1     & 176 & 49  & 9.2 & 52  & 6.3  & 0.99 \\
Emb11W        & 436 & 49  & 14  & 60  & 4.8  & 1.0 \\
SMM1a         & 436 & 38  & 74  & 163 & 8.9  & 13.7 \\
MM18a         & 441 & 37  & 47  & 132 & 9.3  & 9.4 \\
B335          & 165 & 43  & 1.3 & 17  & 4.3  & 0.072 \\ \hline
\end{tabular}
\tablefoot{
\tablefoottext{a}{Adopted distance for the specific source (see discussions and references in Sects.~\ref{firstsource}--\ref{lastsource}).}
\tablefoottext{b}{Bolometric temperature and luminosity from spline fits to the spectral energy distributions in Fig.~\ref{sedfig}.}
\tablefoottext{c}{Radius where the temperature reaches 100~K and column density of H$_2$ within this radius from 1D radiative transfer model fits of a spherical envelope to the spectral energy distribution.}
\tablefoottext{d}{Relative line strengths of a molecular line that is optically thin and arises solely in the part of the envelope where the temperature is above 100~K (assumed to be compact relative to a given beam). Here Emb11W is taken as the reference with a line strength of 1.0.}}
\end{table*}

Based on the SEDs, we also make simple estimates of the extent and mass of warm ($T >$ 100~K) gas for each source. We assume that the gas is distributed in a spherical envelope between 1 and 10,000~au with a density profile $\rho \propto r^{-3/2}$ as expected for free fall. The total luminosity of each source is taken as the heating input. Using the 1D radiative transfer code Transphere\footnote{\url{https://www.ita.uni-heidelberg.de/~dullemond/software/transphere/}} built on the method described in \cite{dullemond02}, we fit the SED at wavelengths between 60~$\mu$m and 1.3~mm to constrain the envelope mass, while dust temperatures are calculated self-consistently. The SED in this wavelength range is expected to be dominated by the bulk envelope emission and the derived temperature profile represents an estimate of average temperatures at given distances from the central stellar heating source (at shorter wavelengths, effects that cannot be accounted for by these 1D models, such as the exact source geometry and scattering, become more important). For the close binaries, the assumption is that they are embedded within a shared envelope and thus that the combined luminosities contribute to the heating of it. From these results, the radius where the temperature exceeds 100~K, as well as the column density of the gas (H$_2$ assuming a standard gas-to-dust mass ratio of 100:1) within this radius are subsequently estimated. In this region, complex organics are expected to sublimate from the icy surfaces of dust grains. Table~\ref{SEDresults} provides these results for each source. We also estimate the expected relative line strengths across the sample (with the model for Emb11W taken as a reference): we assume that the hot gas is unresolved in the observation beam, that the emission is optically thin, and that the excitation temperature of a given species is the same for all sources. Under these assumptions, the line strengths are expected to scale with the total number of molecules within the radius where the temperature reaches above 100~K, i.e., the mass of the gas within this region. The assumption of a common excitation temperature would hold if the presence of the molecule in the gas for all sources is dominated by one chemical process such as the sublimation of the ices at 100~K, placing most molecules in radii where $T\approx 100$~K.

\section{First results}
In this section, we provide a first overview of the data obtained and highlight some key differences that can already be seen from a first qualitative comparison between the emission maps and source spectra.

\subsection{Dust and gas emission maps}
\begin{figure*}
\resizebox{\hsize}{!}{\includegraphics{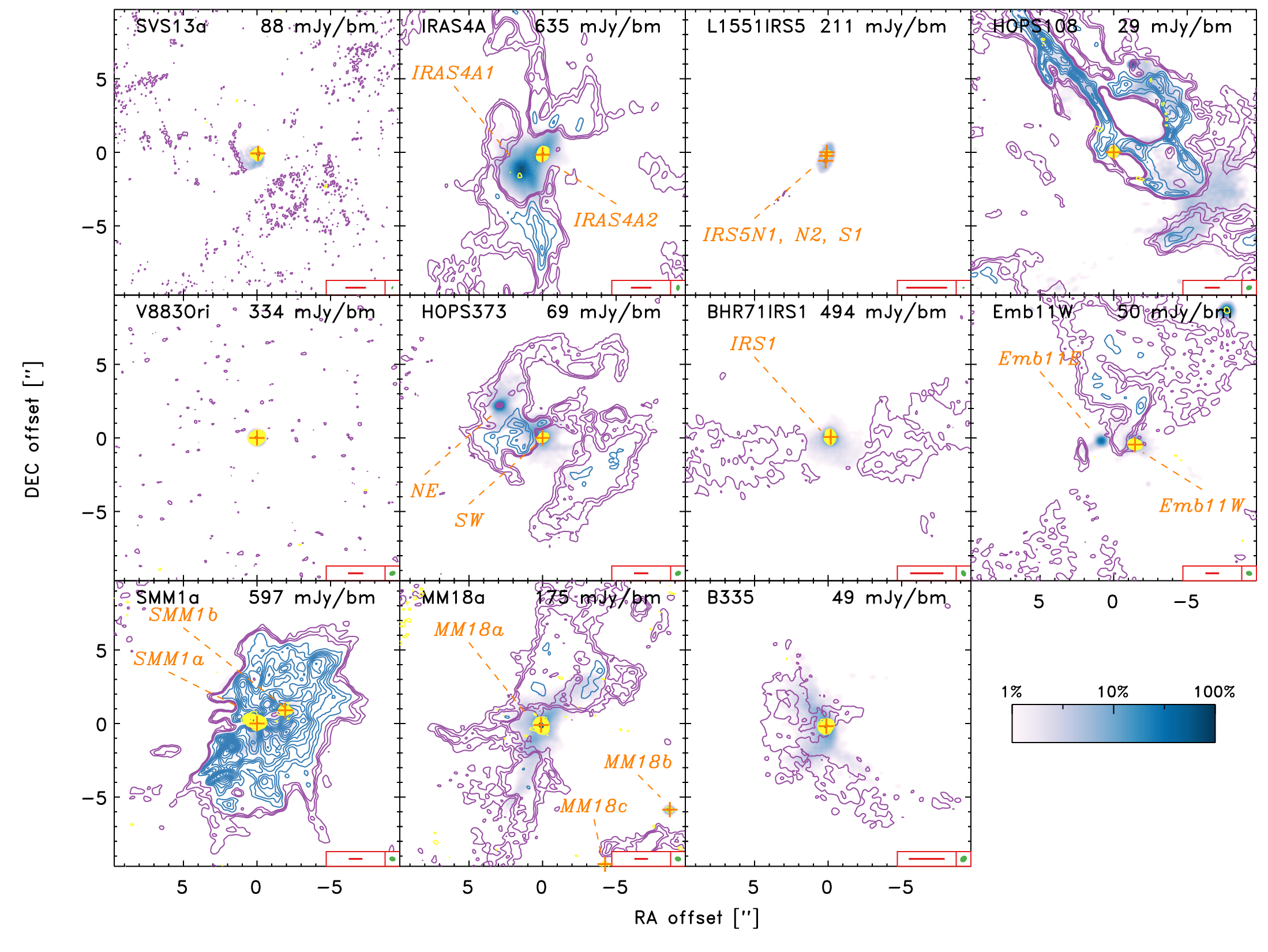}}
\caption{Maps of integrated emission from N$_2$H$^+$ 3--2 (at $3\sigma$, 6$\sigma$, and 9$\sigma$ with purple contours and from there in steps of 6$\sigma$ with blue contours) and $^{13}$CH$_3$OH $5_{1,4}-5_{0,5}$ (in steps of 3$\sigma$ with yellow contours) overlaid on dust continuum images of the 11 targeted sources. The line emission is integrated over $\pm 2$$\sigma_v$ where $\sigma_v$ represents the line width (dispersion) of the lines toward the continuum peaks. The continuum images represent the observed flux density scaled logarithmically from 1 to 100\% of the peak flux density for the individual source given in the top right of each panel and indicated by the color bar in the bottom right of the figure. The beam size and a scale bar corresponding to a linear distance of 400~au are shown in the lower right corner of each panel. The orange plus signs indicate the positions of the peaks of the compact COM emission identified through the channel counting (Table~\ref{t:chancountpos}). The locations of the continuum peaks mentioned in Sect.~\ref{subsec:sample} are indicated with the dashed lines and labeled. The primary beam size at these frequencies is about 20$''$, i.e., matches the sizes of the individual panels.} \label{methanoln2hp}
\end{figure*}

\begin{figure*}
\resizebox{\hsize}{!}{\includegraphics{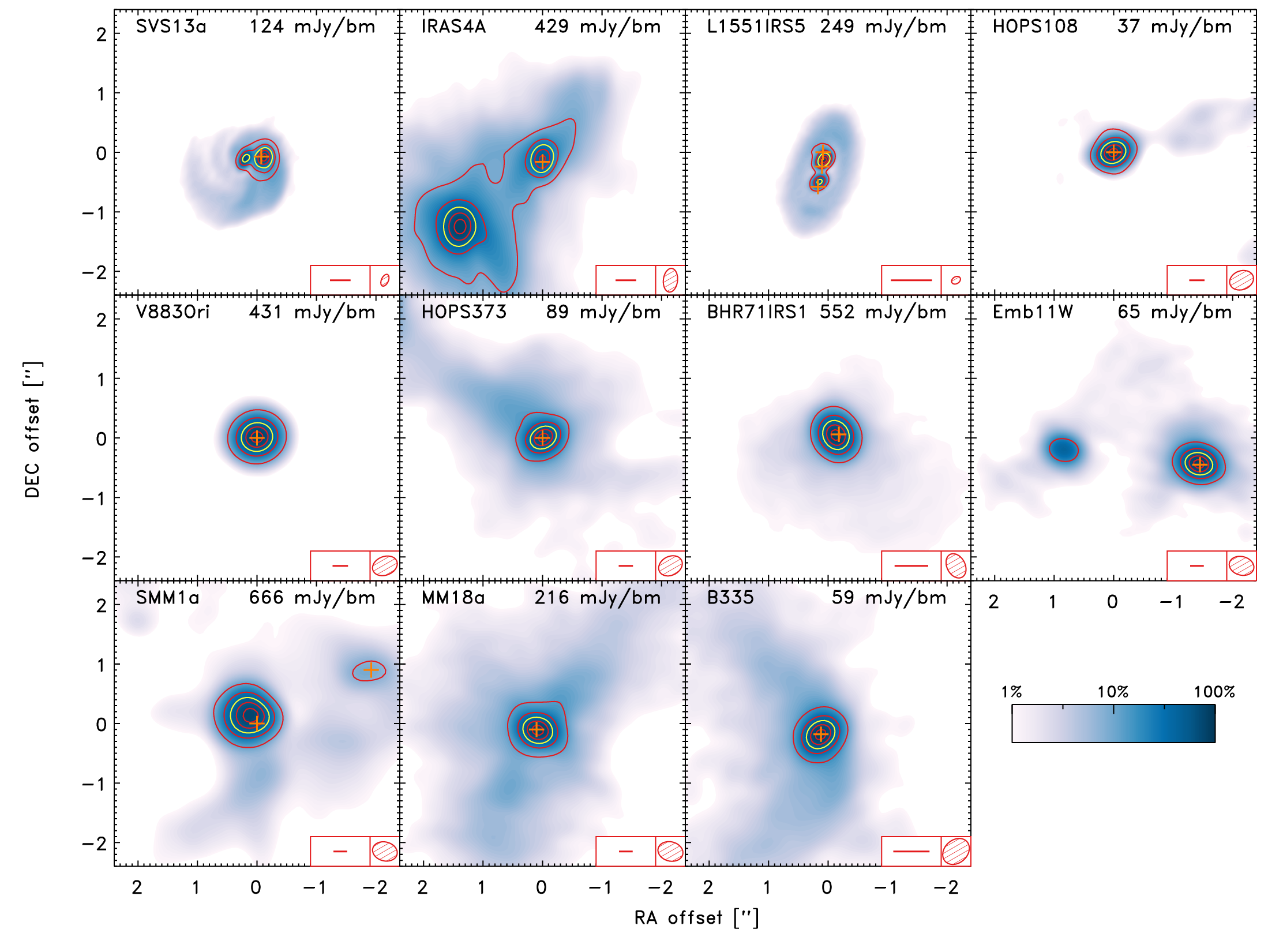}}
\caption{Maps of  emission from the $^{13}$CH$_3$OH $5_{1,4}-5_{0,5}$ transition integrated over the same velocity range as in Fig.~\ref{methanoln2hp} (contours) overlaid on dust continuum images of the 11 targeted sources. The continuum images represent the observed flux density scaled logarithmically from 1 to 100\% of the peak flux density as in Fig.~\ref{methanoln2hp}. The red contours correspond to 10\%, 30\%, 50\%, 70\%, and 90\% of the maximum integrated intensity, while the yellow contour indicates the level of 50\%. The yellow contour for an unresolved source would correspond to the beam full width at half maximum that is shown in the lower right corner of each panel. The red scale-bar in the lower right corner of each panel corresponds to a linear scale of 100~au. As in Fig.~\ref{methanoln2hp}, the orange plus signs indicate the peaks of the compact COM emission from Table~\ref{t:chancountpos}.} \label{methanol13maps}
\end{figure*}

Figure~\ref{methanoln2hp} compares the integrated emission from the N$_2$H$^+$ 3--2 transition ($E_{\rm up} = 27$~K) to that of the $^{13}$CH$_3$OH $5_{1,4}-5_{0,5}$ transition ($E_{\rm up} = 49$~K) at 309.3345~GHz, as well as the dust continuum emission extracted from the lowest-frequency spectral window (279.5~GHz; 1.07~mm), while Fig.~\ref{methanol13maps} zooms in on the $^{13}$CH$_3$OH transition and the continuum emission. The emission from the lines is integrated over $\pm 2\sigma_v$ where $\sigma_v$ represents the velocity dispersion of the lines at the continuum peak. Of these three tracers, the continuum emission is expected to trace the dust in the central circumstellar disks, $^{13}$CH$_3$OH emission from the warm inner envelopes, and N$_2$H$^+$ traces cold material on larger scales. Methanol, CH$_3$OH in particular, is often used as a reference tracer for comparison with other complex organics, and here we chose $^{13}$CH$_3$OH over the main methanol isotopolog because the lines of the latter are often severely optically thick \citep[see also][]{coutens26compass}.

As seen in Fig.~\ref{methanoln2hp}, all sources except the three Class~I sources (L1551IRS5, SVS13a, and V883Ori) are located in highly structured cold environments traced by N$_2$H$^+$ emission. In all cases, the $^{13}$CH$_3$OH emission marks the location of the central protostar(s), which largely reflects the enhanced abundance and emission of the complex organics in the regions heated by the protostars. The $^{13}$CH$_3$OH emission is comparable in size to the beam, highlighting its compact nature. This specific $^{13}$CH$_3$OH transition has an upper energy level of 49~K and, a priori, we would expect that it is predominantly excited in the warmer gas toward the protostar (rather than the more extended, cold envelope material), but could trace material outside the central hot region. Indeed, there is little evidence for extended $^{13}$CH$_3$OH emission in Fig.~\ref{methanol13maps}. The N$_2$H$^+$ emission is deficient in the immediate vicinity of the protostars, similar to what is seen toward several other sources, likely due to sublimation of CO from the icy dust grain mantles at temperatures of 20--30~K, which destroys N$_2$H$^+$ in the gas phase \citep[e.g.,][]{paperii}. The three Class~I sources show mainly compact $^{13}$CH$_3$OH emission, again centered on protostars, and overall very little N$_2$H$^+$ emission. This likely reflects the fact that their cold envelopes ($T < 20$~K) have largely dissipated, leaving only warm, compact emission.

The continuum maps of Fig.~\ref{methanol13maps} show significant diversity, with varying degrees of extended emission and substructure. Known strong continuum sources such as IRAS4A1 and Serpens-Emb11E are seen within the fields of view of IRAS4A and Emb11W, respectively, and the tight binaries in SVS13a and L1551IRS5 are resolved in the data from the more extended configurations obtained for those two sources. Several other sources show additional continuum components within the ALMA primary beam, including SMM1a, HOPS108, and HOPS373. The more deeply embedded sources typically show extended features that reflect high dust densities or elevated temperatures. Some structures appear related to the presence of outflow cavities (e.g., B335), while others are more complex (e.g., SMM1a). For L1551IRS5, the two individual disks around each individual component of the binary are seen, as well as the more extended, well separated circumbinary disk. SVS13a shows the prominent spiral structures in continuum emission that originate in the dusty circumbinary disk previously reported \cite[][]{tobin18,diazrodriguez22}.

To search systematically for hot corino-like emission (i.e., compact emission of COMs) in each field of view of the COMPASS survey, we applied the method of
\cite{belloche19,belloche20} to build maps of channel counts. Appendix~\ref{channelcounts} describes the analysis in more detail but, in short, the number of channels with emission flux above a certain level are counted for each spatial pixel of the data cube. The peaks of the line counts in the resulting maps are listed in Table~\ref{t:chancountpos} and are included as orange plus signs in Figs.~\ref{methanoln2hp} and \ref{methanol13maps}. 
\begin{table*}[!t]
 \begin{center}
 \caption{
 Coordinates and systemic velocity of channel count peaks present in the field of view of each COMPASS source.
}
 \label{t:chancountpos}
 \vspace*{-1.2ex}
 \begin{tabular}{lrrrrrl}
 \hline\hline
 \multicolumn{1}{c}{Field} & \multicolumn{2}{c}{Coordinates\tablefootmark{a}} & \multicolumn{2}{c}{Offsets\tablefootmark{b}} & \multicolumn{1}{c}{$V_\mathrm{lsr}$\tablefootmark{c}} & Source name \\ 
  & \multicolumn{1}{c}{hh:mm:ss.sss} & \multicolumn{1}{c}{dd:mm:ss.ss} & \multicolumn{1}{c}{[$\arcsec$]} & \multicolumn{1}{c}{[$\arcsec$]} & \multicolumn{1}{c}{[km s$^{-1}$]} & \\ 
 \hline
SVS13a & 03:29:03.750 & $31$:16:03.72 & $-0.08$ & $-0.08$ & $7.7$ & SVS13 VLA4A
 \\ 
\hline 
IRAS4A & 03:29:10.433 & $31$:13:31.97 & $ 0.01$ & $-0.15$ & $6.8$ & IRAS4A2
 \\ 
\hline 
L1551IRS5 & 04:31:34.166 & $18$:08:04.72 & $ 0.08$ & $ 0.00$ & $9.7$ & L1551 IRS5N1
 \\ 
 & 04:31:34.167 & $18$:08:04.48 & $ 0.11$ & $-0.24$ & $5.9$ & L1551 IRS5N2
 \\ 
 & 04:31:34.172 & $18$:08:04.14 & $ 0.18$ & $-0.59$ & $3.1$ & L1551 IRS5S1
 \\ 
\hline 
HOPS108 & 05:35:27.085 & $-05$:10:00.08 & $ 0.02$ & $-0.01$ & $12.2$ & HOPS108
 \\ 
\hline 
V883Ori & 05:38:18.101 & $-07$:02:25.99 & $ 0.01$ & $ 0.01$ & $4.1$ & V883 Ori
 \\ 
\hline 
HOPS373 & 05:46:30.904 & $-00$:02:35.19 & $-0.02$ & $ 0.02$ & $9.9$ & HOPS373 SW
 \\ 
\hline 
BHR71IRS1 & 12:01:36.470 & $-65$:08:49.33 & $-0.18$ & $ 0.05$ & $-4.6$ & BHR71 IRS1
 \\ 
 & 12:01:34.008 & $-65$:08:48.07 & $-15.70$ & $ 1.31$ & $-4.5$ & BHR71 IRS2
 \\ 
\hline 
Emb11W & 18:29:06.624 & $00$:30:33.86 & $-1.44$ & $-0.44$ & $8.5$ & Ser-emb-11W
 \\ 
\hline 
SMM1a & 18:29:49.794 & $01$:15:20.21 & $ 0.01$ & $ 0.01$ & $7.5$ & Serp SMM1a
 \\ 
 & 18:29:49.666 & $01$:15:21.09 & $-1.91$ & $ 0.89$ & $8.1$ & Serp SMM1b
 \\ 
\hline 
MM18a & 18:30:04.124 & $-02$:03:02.66 & $ 0.09$ & $-0.11$ & $7.9$ & SerpS MM18a
 \\ 
 & 18:30:03.541 & $-02$:03:08.39 & $-8.66$ & $-5.84$ & $7.3$ & SerpS MM18b
 \\ 
 & 18:30:03.835 & $-02$:03:12.10 & $-4.24$ & $-9.55$ & $9.3$ & SerpS MM18c
 \\ 
\hline 
B335 & 19:37:00.902 & $07$:34:09.41 & $ 0.13$ & $-0.18$ & $6.9$ & B335
 \\ 
\hline 
 \hline
 \end{tabular}
 \end{center}
 \vspace*{-2.5ex}
 \tablefoot{
 \tablefoottext{a}{Equatorial coordinates (ICRS).}
 \tablefoottext{b}{Equatorial offsets relative to the phase center.}
 \tablefoottext{c}{Systemic velocity derived from $^{13}$CH$_3$OH.}
 }
 \end{table*}

\subsection{Spectra and line densities}\label{overviewspectra}
Figure~\ref{bhr71fullspec} shows the full spectrum toward BHR71-IRS1 extracted at the peak from the line counting process (which for this source coincides with the position of the continuum emission peak), and Fig.~\ref{bhr71fullspec_zoom} displays a version zoomed in on the intensity scale to illustrate the presence of faint transitions. The locations of strong lines of a range of simple species are indicated. Although several prominent emission peaks are associated with these, it is also clear that there is significant emission from other species, including complex organics, at similar line strengths (above 100 mJy~beam$^{-1}$ peak intensities). Some species such as N$_2$H$^+$, N$_2$D$^+$, DCN, and DNC are not seen in emission toward this position, but rather in absorption, reflecting the combined effects of absorption against  compact continuum emission and resolving out of extended line emission due to the lack of short spacings. The broad redshifted and blueshifted emission offset from the systemic velocity in the SiO 7--6 transition at 303.9268~GHz is a result of shocks related to the outflow driven by this protostar.

The high line density is further emphasized by the zoom-in spectrum in Fig.~\ref{bhr71fullspec_zoom}. At this level, the forest of significant, but fainter lines is mainly attributed to complex organics and their isotopologs. It is worth highlighting the amount of information in just one single spectrum extracted for one position from the field of view of observations that, in addition to the scientific potential, present the challenge of handling a volume of image data across $\sim$ 138,000 channels that must be imaged, stored, and interpreted.

\begin{figure*}
\resizebox{\hsize}{!}{\includegraphics{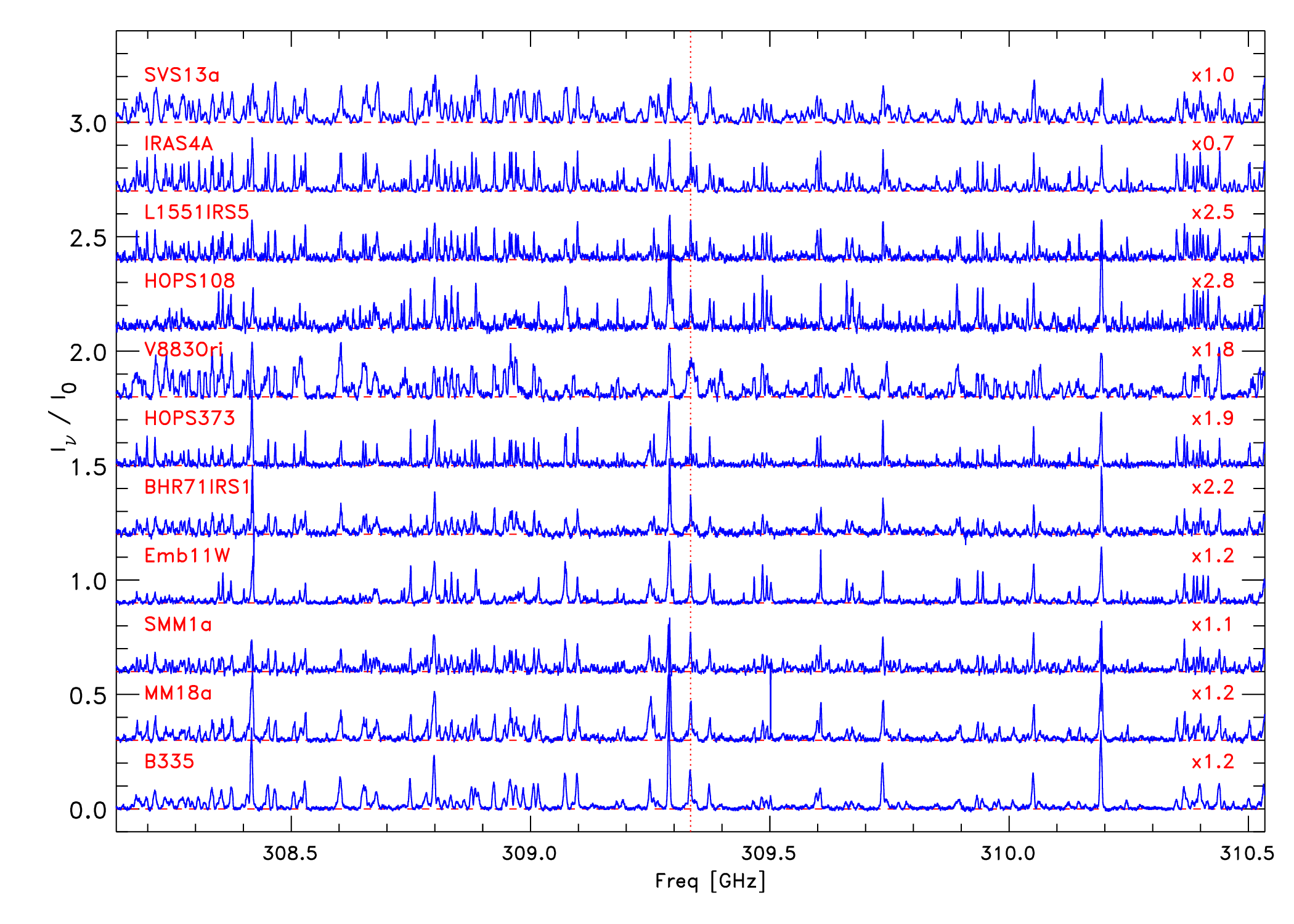}}
\caption{Comparison between the spectra for the 11 targeted sources zoomed in on a 2.4~GHz frequency window. All spectra are normalized so that the peak intensity of the $^{13}$CH$_3$OH $5_{1,4}-5_{0,5}$ transition at 309.3345~GHz (vertical dotted red line) matches that of that specific transition observed toward SVS13a. The factor by which the individual spectra have been scaled is given in the right hand side of each.}\label{spectracomparisons}
\end{figure*}

Figure~\ref{spectracomparisons} compares the 11 targeted sources over a 2.4~GHz frequency window centered on the $^{13}$CH$_3$OH $5_{1,4}-5_{0,5}$ transition at 309.3345~GHz and scaled to the peak intensity of SVS13a for this transition (0.18~Jy~beam$^{-1}$). The absolute scale varies by about a factor of 3.5 from 0.062~Jy~beam$^{-1}$ (HOPS108) to 0.21~Jy~beam$^{-1}$ (IRAS4A). As expected from the sample selection, all sources show line-rich spectra already at this scale, where the brightest lines are about two orders of magnitude above the RMS level per channel (i.e., the noise level is not discernible in these plots). However, variations in the relative line intensities from source to source are already seen, suggesting the need for a careful analysis of the excitation temperatures of individual species and their relative abundances. Applying automated line counts for each source, we find that the number of lines above 5$\sigma$ is $\approx$5000--9000 per source for the 32.7~GHz bandwidth of COMPASS, i.e., corresponding to $\approx$4--8~km~s$^{-1}$ bandwidth per counted line. Given that most sources have line widths of 2.5--5.0~km~s$^{-1}$ full width at half maximum (FWHM), this implies that some portions of the spectra will be line-confused, further emphasizing the importance of careful spectral modeling for the identification of faint lines and rare species.

\subsection{First-order comparison of observed and predicted line strengths}
\begin{figure*}
\resizebox{\hsize}{!}{\includegraphics{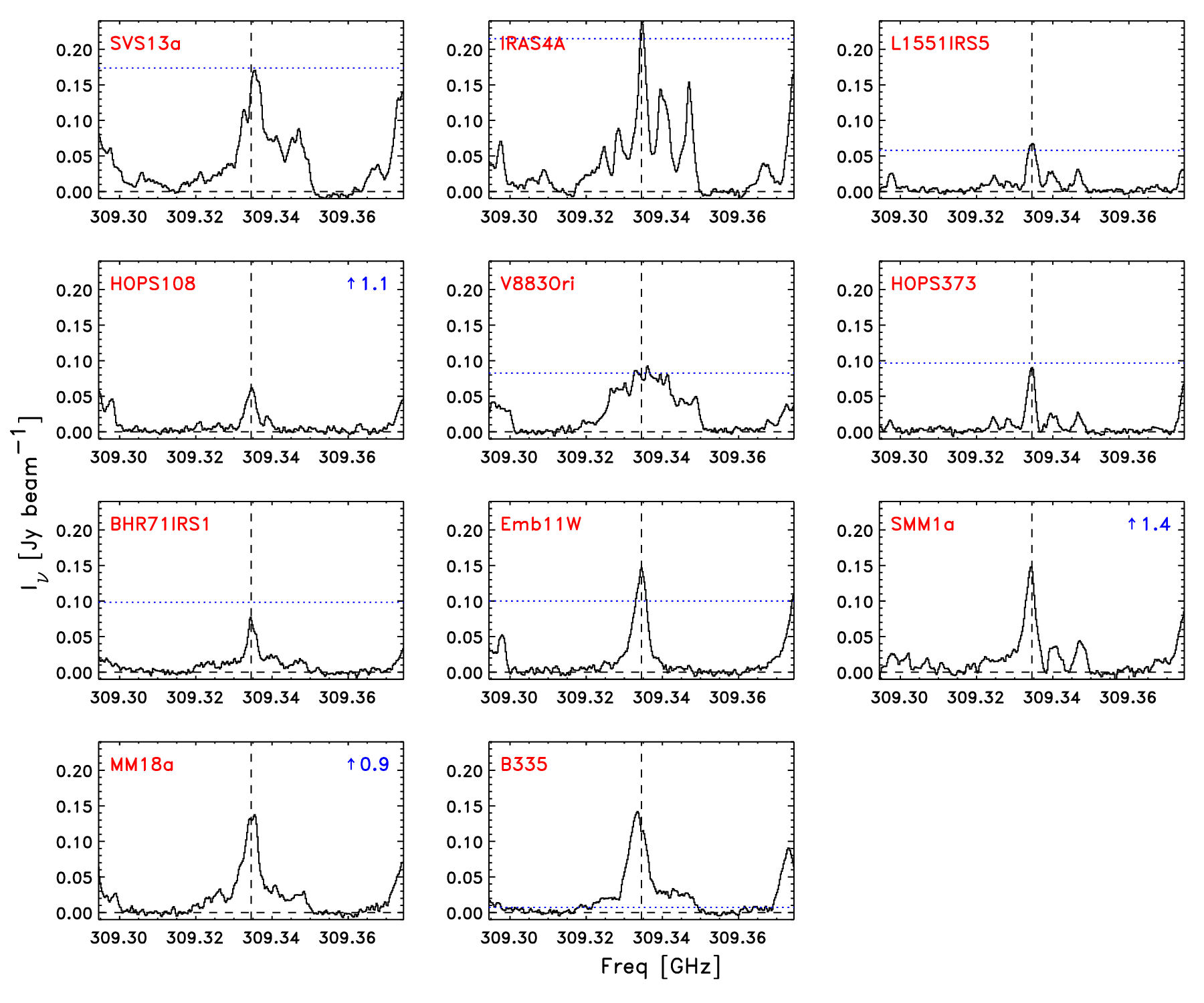}}
\caption{Spectrum of the $^{13}$CH$_3$OH $5_{1,4}-5_{0,5}$ line (indicated with the vertical dashed line) toward the peak of emission seen in Fig.~\ref{methanol13maps}. For each source, we show the predicted strength of the line emission assuming that it is solely determined by the mass of warm  $T > 100$~K gas from the 1D radiative transfer models and scaled so that the modeled line peak flux density is 0.1~Jy~beam$^{-1}$ for Emb11W. For three sources, HOPS108, SMM1a, and MM18a, the predicted intensity is higher than what is shown in this figure: for these sources, the predicted strength is given in the top right of the respective panel.}\label{methanol13spectrum}
\end{figure*}

Figure~\ref{methanol13spectrum} shows the spectra of the $^{13}$CH$_3$OH $5_{1,4}-5_{0,5}$ transition (309.3345~GHz) extracted from the peak emission positions from the channel counting described above. For each spectrum, the horizontal dotted line indicates the relative predicted strength of this transition toward the different sources assuming that the emission originates in the warm gas ($T > 100$~K) of each hot corino with the extent and peak column density, or equivalent total number of molecules or mass of envelope material with temperatures higher than 100~K, constrained by the 1D radiative transfer model fits to the SEDs and under the assumption of a common excitation temperature for all sources, as described in Sect.~\ref{seds}. The dotted lines are scaled so that the total number of $^{13}$CH$_3$OH molecules in the warm gas toward Emb11W would correspond to a peak line strength of 0.10~Jy~beam$^{-1}$ (i.e., the emission is assumed to be optically thin, so the line flux directly traces the total number of ``warm'' molecules).

The comparisons show good agreement for many sources, suggesting that the scaling by luminosity and overall mass is a reasonable first-order proxy for the overall column density of the warm gas. This also implies that $^{13}$CH$_3$OH emission serves as an effective reference tracer to estimate the abundances of other species. There are some exceptions, namely HOPS108, SMM1a, and MM18a, where much stronger lines are predicted: those are the most luminous and massive in the sample, and the mismatch is likely a result of the emission being more extended than the beam and possibly the column densities high enough that the $^{13}$CH$_3$OH transition becomes optically thick. The opposite situation presents itself for B335, for which a much weaker transition is predicted. However, it should be noted that that source has been found to have undergone a recent luminosity outburst –– and the stronger COM emission for that source may therefore reflect the more extended warm gas of the post-burst state \citep{lee25}.

\subsection{Empirical comparison of relative contributions from species}\label{empiricalcomparisons}
Figures~C.1--C.15 in Appendix~C\footnote{\url{https://zenodo.org/records/19843377}} show the 15 lines predicted to be brightest for seven key species for each of the 11 sources in the sample. The seven species consist of five representative complex organics observed toward solar-type protostars ($^{13}$CH$_3$OH as a proxy for the main methanol isotopolog, CH$_3$CN, CH$_3$OCHO, CH$_3$CHO, and NH$_2$CHO), as well as CH$_2$DOH as a tracer of deuteration of complex organics, and HNCO as a key species related to nitrogen chemistry. The decision about which lines to include was made by calculating synthetic spectra for each species assuming an excitation temperature of 150~K under the assumption that the emission of all lines is optically thin. The lines shown are thus the same for all 11 sources. From the list of lines predicted to be among the 15 brightest, we removed one line from each for methyl formate (CH$_3$OCHO) and formamide (NH$_2$CHO) that were blended with other brighter transitions from other species. In addition, in cases where multiple transitions are merged into one spectral feature, only the composite feature is shown.

\begin{figure*}
\resizebox{\hsize}{!}{\includegraphics{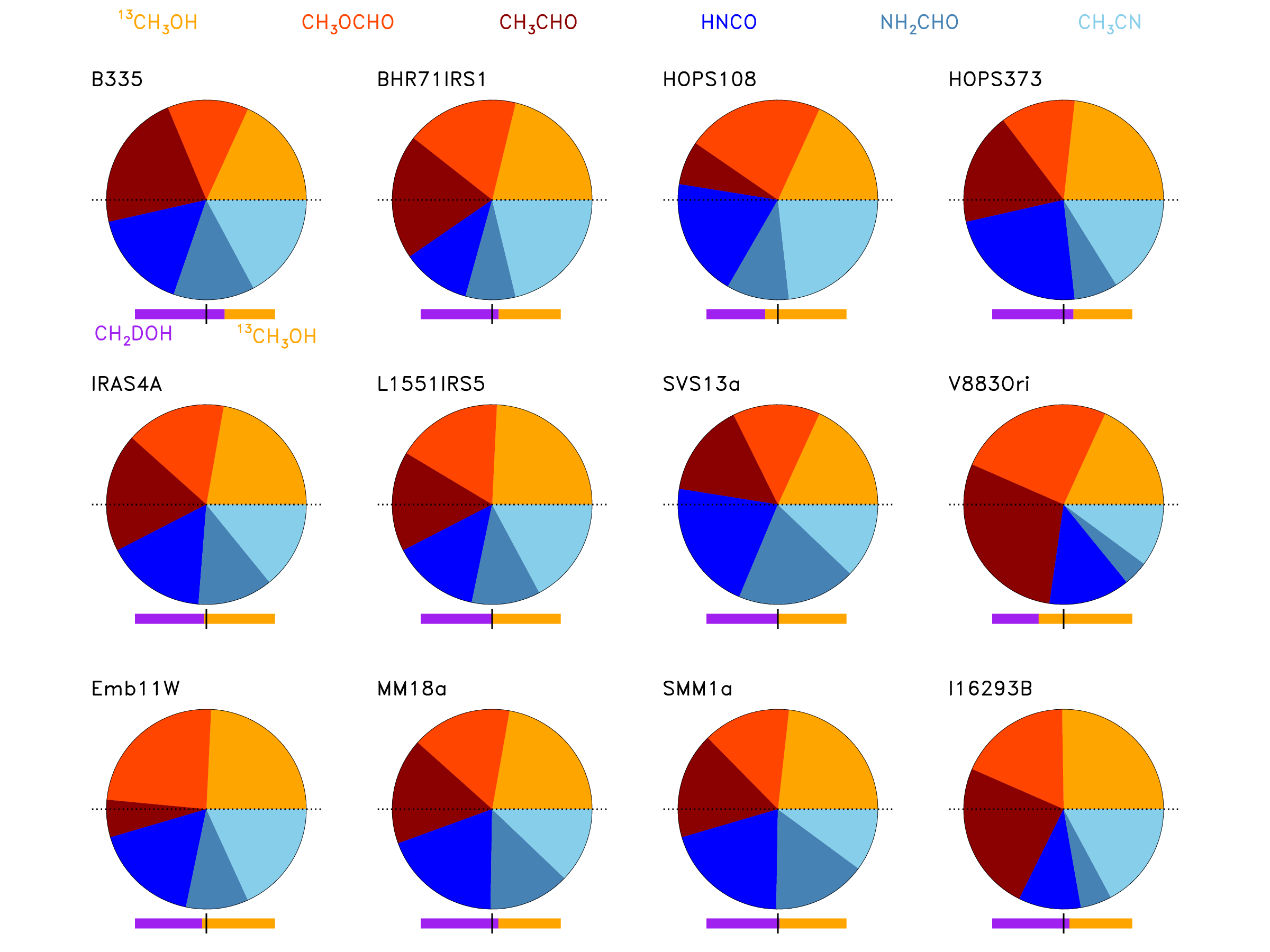}}
\caption{Fraction of the total line integrated intensities supplied by the fifteen lines predicted to be the brightest for key organic species for each of the COMPASS sources, as well as (model-predicted) IRAS16293B. The pie charts represent the relative contributions from $^{13}$CH$_3$OH, CH$_3$OCHO, CH$_3$CHO, HNCO, NH$_2$CHO, and CH$_3$CN (see legend at the top of the figure) to the total integrated intensities from the lines of all six species, while the bar in the bottom compares the relative fractions from $^{13}$CH$_3$OH and CH$_2$DOH compared to the total integrated intensities of the lines of these two species together. The fifteen measured lines (also included in Appendix~C) are the ones predicted to be the strongest assuming a temperature of 150~K and optically thin emission, excluding one transition each from CH$_3$OCHO and NH$_2$CHO that are clearly blended with other strong lines. It is thus the same set of lines that are compared for all sources. For IRAS16293B, the line strengths are taken from predictions made with synthetic spectra based on the PILS abundances extended to the same frequency range as covered by COMPASS.}\label{piechart}
\end{figure*}

The spectra reveal notable differences between the sources, for example, with the nitrogen-bearing species showing variations from source to source in comparison to some of the other species. For a first empirical estimate of these variations, we calculated the line intensity integrated over $\pm 1$~km~s$^{-1}$ around the rest frequencies of the lines shifted according to the local standard rest (LSR) velocities of the individual sources as a systematic and quantitative estimate of the relative line strengths across the sample. If the emission is optically thin and there are no substantial variations in terms of velocity offset or line widths between different species, or transitions for one species, these estimates allow us to measure and compare the relative luminosities of the different species for the individual sources directly. Figure~\ref{piechart} shows the relative flux emitted by each species estimated in this manner for the different sources: in the pie charts, the fraction of emission from the key oxygen-bearing species ($^{13}$CH$_3$OH, CH$_3$OCHO, and CH$_3$CHO) is compared with those of the key nitrogen-bearing species (CH$_3$CN, HNCO, and NH$_2$CHO). It is important to note that while this approach gives first impressions of possible trends and the expected level of variations, it does not factor in possible differences in excitation conditions or line opacities in different sources or for specific species. In addition to the pie charts, the figure shows a bar for each source representing the relative contributions of the predicted brightest lines of $^{13}$CH$_3$OH and CH$_2$DOH compared to the total flux of the brightest lines of these two species. This bar visualizes their relative luminosities as a proxy for their relative abundances, i.e., the methanol deuteration, although with the same caveats noted above.

To allow a comparison with IRAS16293B, we created synthetic spectra of the same species for that source, adopting column densities and temperatures measured as part of PILS \citep[see, e.g., compilation of column densities in][]{jorgensen20} for the spectral range observed in COMPASS. The synthetic spectra take into account the possible optical thickness of the lines. Thus, these are a direct prediction for what the observed spectra would be for IRAS16293B over the spectral range of COMPASS (including both optically thin and thick transitions).

A comparison of the pie charts of Fig.~\ref{piechart} highlights variations in the oxygen and nitrogen content across the sample with about half of the line flux coming from the three nitrogen-bearing species for B335, HOPS108, and SVS13a, for example, but only about a quarter for V883Ori and IRAS16293B. For these two sources, in particular, NH$_2$CHO is not contributing much to the total line emission budget: the lower line flux of nitrogen-bearing organics toward V883Ori is in line with the recent results of \cite{jeong25v883ori}, who indeed inferred a low abundance of these species toward this source. Similarly, a lower nitrogen content for IRAS16293B compared to the Galactic Center source SgrB2(N2) and comet 67P/Churyumov–Gerasimenko was also mentioned in \cite{jorgensen20}. Several explanations for variations in the abundance of complex organic molecules bearing nitrogen and oxygen have been proposed related to physical mechanisms that lead to sublimation of species from the ices or their formation in the gas phase, including the importance of accretion shocks \citep[e.g.,][]{csengeri19}; the destruction of carbonaceous grains during accretion bursts \citep[e.g.,][]{vanthoff20}; the balance between the formation of parent species in the ices and the subsequent production or destruction in the gas phase \citep[e.g.,][]{nazari21}, or the degree to which the ice mantles are desorbed \citep{belloche25}. Another difference is seen between the relative contributions to the line fluxes from CH$_3$OCHO and CH$_3$CHO: CH$_3$OCHO is clearly the more prominent of these two species toward Emb11W and HOPS108, but much less so toward, for example, HOPS373, V883Ori, and B335. Finally, the comparison between $^{13}$CH$_3$OH and CH$_2$DOH suggests that variations in D/H ratios are expected to be found across the sample with, for example, V883Ori showing a CH$_2$DOH/$^{13}$CH$_3$OH line luminosity ratio below unity, but other sources such as B335 and I16293B yielding ratios above 1.

Of these three selected comparisons, it is noteworthy that there is no obvious correlation or anti-correlation, for example, between the O/N line flux ratio and the level of deuteration. There is also no clear trend that sources in specific star-forming environments, for example, systematically show any peculiar signatures. Overall, the differences in the relative line flux contributions are no larger than factors of 5--8 even in extreme cases. However, because the comparison is purely empirical, based on measured line fluxes without accounting for excitation or optical-depth effects, robust conclusions about abundance variations will require detailed spectral modeling of all transitions across the full COMPASS bandwidth for each species and each source. This will be addressed in future COMPASS papers.

\section{Conclusions}
In this work, we have outlined the motivation and strategy behind the ALMA Large Program COMPASS and presented initial insights from its data. The program seeks to determine how the chemical inventories of protostars vary with environmental and evolutionary factors, with a particular focus on complex organic molecules. This is achieved through unbiased ALMA Band 7 spectral surveys (279.0--311.7 GHz) of 11 protostellar sources previously identified as hosting warm gas ($T > 100$~K) enriched in complex organics.

Preliminary maps confirm that these molecules are concentrated in warm, compact regions. Comparisons to predictions from 1D radiative transfer models support a picture where to first order the extent of these regions is set by the sublimation of molecules at temperatures above 100~K, which in turn is regulated by a combination of the source luminosities and dust mass. An initial empirical characterization of key organics based on their relative line luminosities suggests the presence of source-to-source chemical differences. However, detailed spectral modeling will be required to quantify abundance variations below an order of magnitude.

Looking ahead, COMPASS aims to systematically establish the molecular inventories of these protostars and explore potential isotopic differences among them. Ultimately, the program seeks to determine whether the chemical complexity of young stars -- and by extension protoplanetary disks and emerging planets -- is primarily shaped by evolutionary history or by the local environment in which they form.

\section*{Data availability}
Appendix~C is available at \url{https://zenodo.org/records/19843377}.

\begin{acknowledgements}
The authors are deeply grateful to everyone at the ALMA observatory that enabled the successful execution of this Large Program and, in particular, staff at the Nordic ALMA Regional Centre node for their support during its planning and execution stages. This paper makes use of the following ALMA data: ADS/JAO.ALMA\#2022.1.00316.L. ALMA is a partnership of ESO (representing its member states), NSF (USA) and NINS (Japan), together with NRC (Canada), NSTC and ASIAA (Taiwan), and KASI (Republic of Korea), in cooperation with the Republic of Chile. The Joint ALMA Observatory is operated by ESO, AUI/NRAO and NAOJ. The National Radio Astronomy Observatory is a facility of the National Science Foundation operated under cooperative agreement by Associated Universities, Inc. The project that led to this publication has received support from ORP that is funded by the European Union’s Horizon 2020 research and innovation programme under grant agreement No 101004719 [ORP]. This research has made use of the NASA/IPAC Infrared Science Archive, which is funded by the National Aeronautics and Space Administration and operated by the California Institute of Technology. It also made use of data from the \emph{Herschel} Gould Belt Survey (HGBS) project (\url{http://gouldbelt-herschel.cea.fr}). The HGBS is a Herschel Key Programme jointly carried out by SPIRE Specialist Astronomy Group 3 (SAG 3), scientists of several institutes in the PACS Consortium (CEA Saclay, INAF-IFSI Rome and INAF-Arcetri, KU Leuven, MPIA Heidelberg), and scientists of the Herschel Science Center.

We acknowledge financial support from the following: the European Research Council (ERC) under the European Union’s Horizon 2020 research and innovation programme (ERC Starting Grant ``Chemtrip'', grant agreement No 949278; \emph{A.C., F.C.-S.M., A.A. and P.M.}); the National Research Foundation of Korea (NRF), funded by the Ministry of Science and ICT (MSIT) under grant numbers 2021R1A2C1011718 and RS-2024-00416859 (\emph{J.-E.L., J.-H.J., H.-S.Y., and C.-H.K.}), and by the Ministry of Education through the Basic Science Research Program under grant number RS-2025-25402622 (\emph{J.-H.J.}); the Thematic Action ``Physique et Chimie du Milieu Interstellaire'' (PCMI) of INSU Programme National ``Astro'', with contributions from CNRS Physique \& CNRS Chimie, CEA, and CNES (\emph{S.M.}); the Ministry of Education of Taiwan (a grant from Center for Informatics and Computation in Astronomy, CICA,  and grant number 110J0353I9) and the National Science and Technology Council, Taiwan (Grant NSTC111-2112-M-007-014-MY3, NSTC113-2639-M-A49-002-ASP, and NSTC113-2112-M-007-027; \emph{D.H.}); the National Radio Astronomy Observatory and the North American ALMA Science Center (\emph{Z.T. and M.R.}); the NKFIH through NKKP grant ADVANCED 149943 and Excellence grant TKP2021-NKTA-64 (project no.149943 has been implemented with the support provided by the Ministry of Culture and Innovation of Hungary from the National Research, Development and Innovation Fund, financed under the NKKP ADVANCED funding scheme) as well as COST (European Cooperation in Science and Technology) through COST Action PLANETS CA22133 (A.K.); the Max Planck Society (\emph{S.S.}); the ESO and IAU Gruber Foundation Fellowship programs (\emph{P.N.}); the Ministry of Education, Culture, Sports, Science, and Technology of Japan (``Grant-in-Aid''; 20H05845, 20H05844) and RIKEN (pioneering project ``Evolution of Matter in the Universe''; \emph{(Y.-L.Y.)}; RIKEN through the Special Postdoctoral Researcher Program (Fellowships; \emph{J.F.A. and S.Z.}).
\end{acknowledgements}

\appendix
\section{Channel count maps}\label{channelcounts}
In this section we describe the method for identifying positions with compact emission of COMs for further analysis. The approach follows the method of \citet{belloche19,belloche20} where the number of channels with emission above a certain threshold is counted for each spatial pixel in the continuum-subtracted datacubes. In the analysis here we counted the 
number of channels that have a flux density higher than 6 times the rms noise 
level. The rms noise level was estimated with the task ``go noise'' of the GREG 
software\footnote{See http://www.iram.fr/IRAMFR/GILDAS}, which computes the 
median of the dispersions of the intensity distributions of all channel maps of 
a cube. Before counting the channels, we masked frequency ranges that potentially could include 
emission from simple molecules or low-energy transitions of methanol. These 
frequency masks are listed in Table~\ref{t:freqmasks}. The width of each mask 
was chosen such that it covers the velocity ranges of the potential emission of 
all compact sources in all fields of view of the COMPASS survey. Several masks 
are much broader than others (for instance for SO or SiO) to avoid 
high-velocity emission related to outflows. Three masks were set around ozone lines because these atmospheric features slightly increase the noise level in these frequency ranges. In total, the masks cover 4.1 GHz, which represents about 13\% of the frequency range covered by the COMPASS survey.

\begin{figure}[!b]
\centerline{\resizebox{\hsize}{!}{\includegraphics[angle=0]{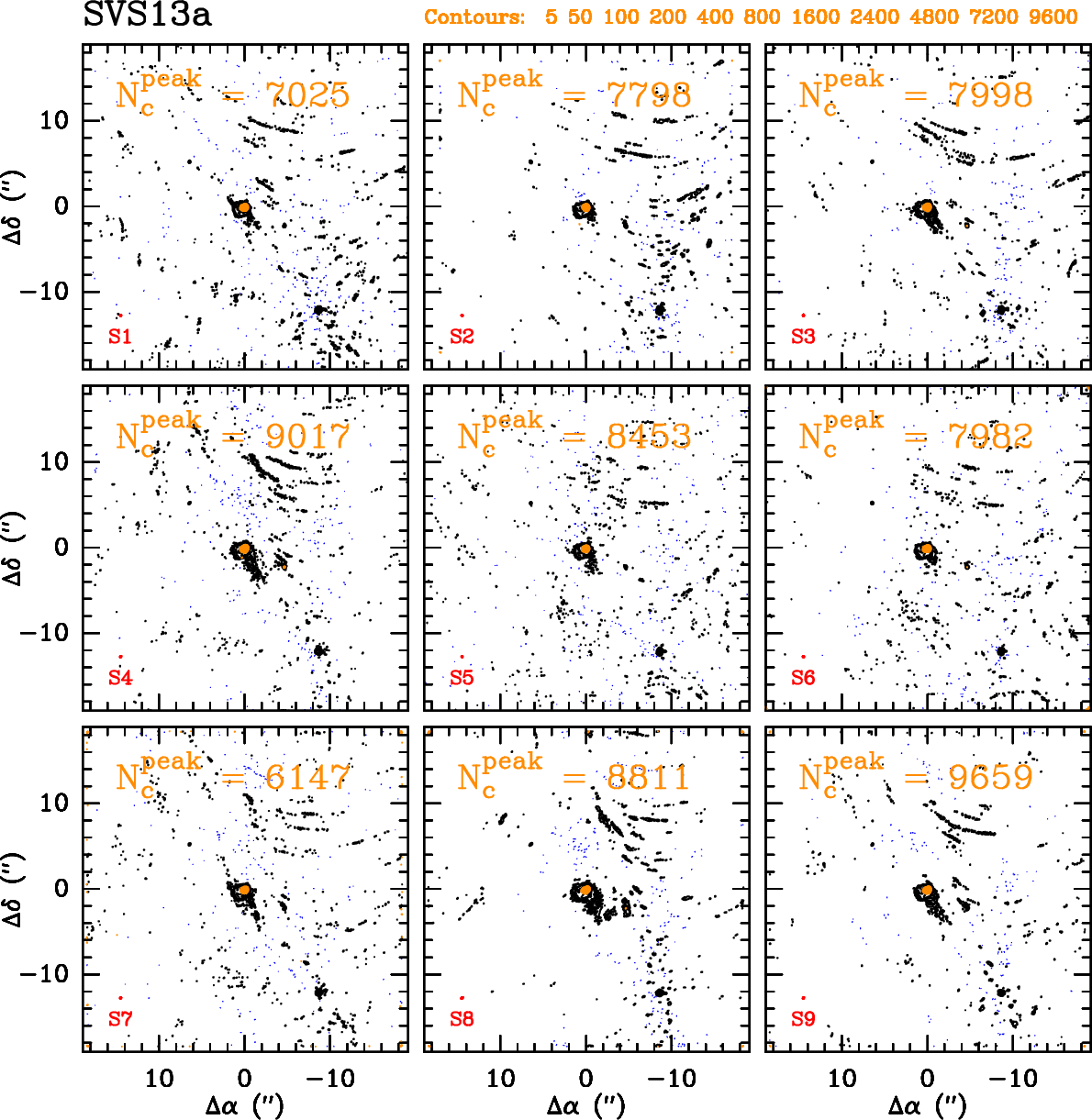}}}
\caption{Contour maps of the number of channels with continuum-subtracted flux
density above the 6$\sigma$ level for each of the nine COMPASS spectral setups 
observed toward SVS13A (orange contours). The contour levels are indicated at 
the top. The channel count peak is indicated in each panel and the small red ellipse (above the setting number, e.g., ``S1'')
represents the beam size (HPBW). The black contours represent the continuum 
emission. They start at a signal-to-noise ratio of 3 and increase by a factor of 2 at each step.
The blue dashed contours represent the negative levels. The equatorial offsets 
are relative to the phase center.}
\label{f:chancount_svs13a}
\end{figure}

\begin{figure}[!ht]
\centerline{\resizebox{\hsize}{!}{\includegraphics[angle=0]{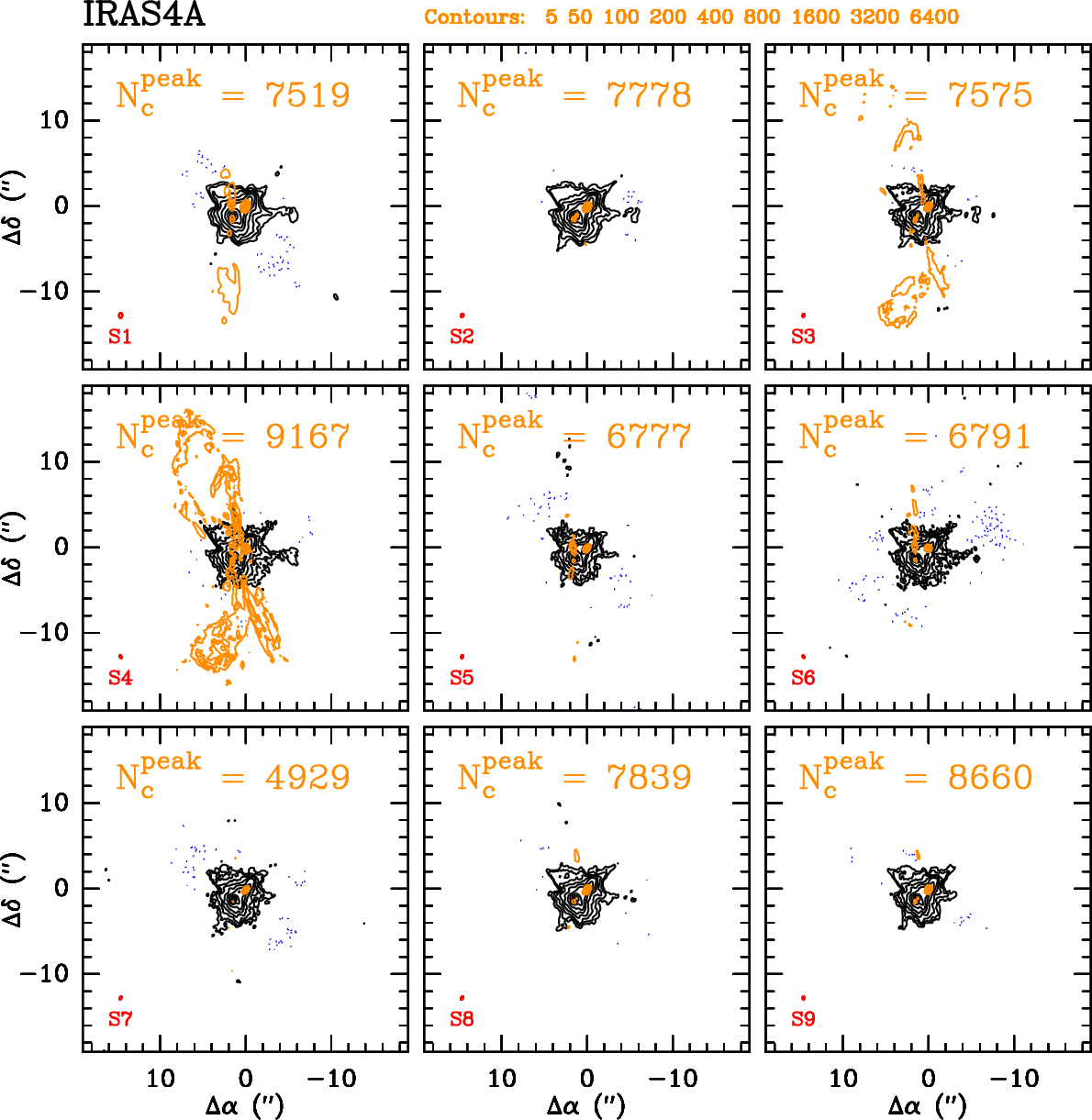}}}
\caption{Same as Fig.~\ref{f:chancount_svs13a}, but for IRAS4A.}
\label{f:chancount_iras4a}
\end{figure}

\begin{figure}[!ht]
\centerline{\resizebox{\hsize}{!}{\includegraphics[angle=0]{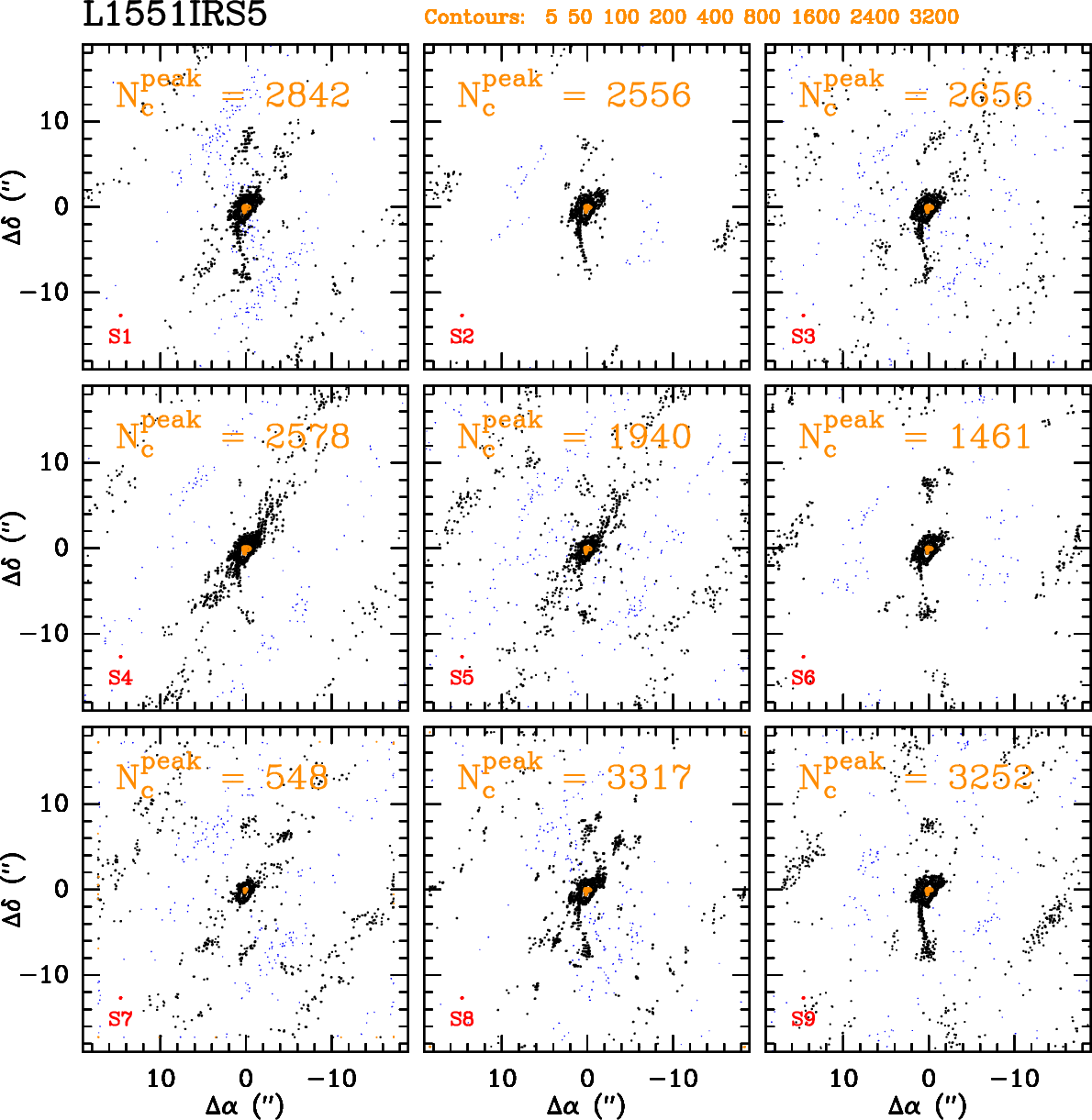}}}
\caption{Same as Fig.~\ref{f:chancount_svs13a}, but for L1551IRS5.
}
\label{f:chancount_l1551irs5}
\end{figure}

\begin{figure}[!ht]
\centerline{\resizebox{\hsize}{!}{\includegraphics[angle=0]{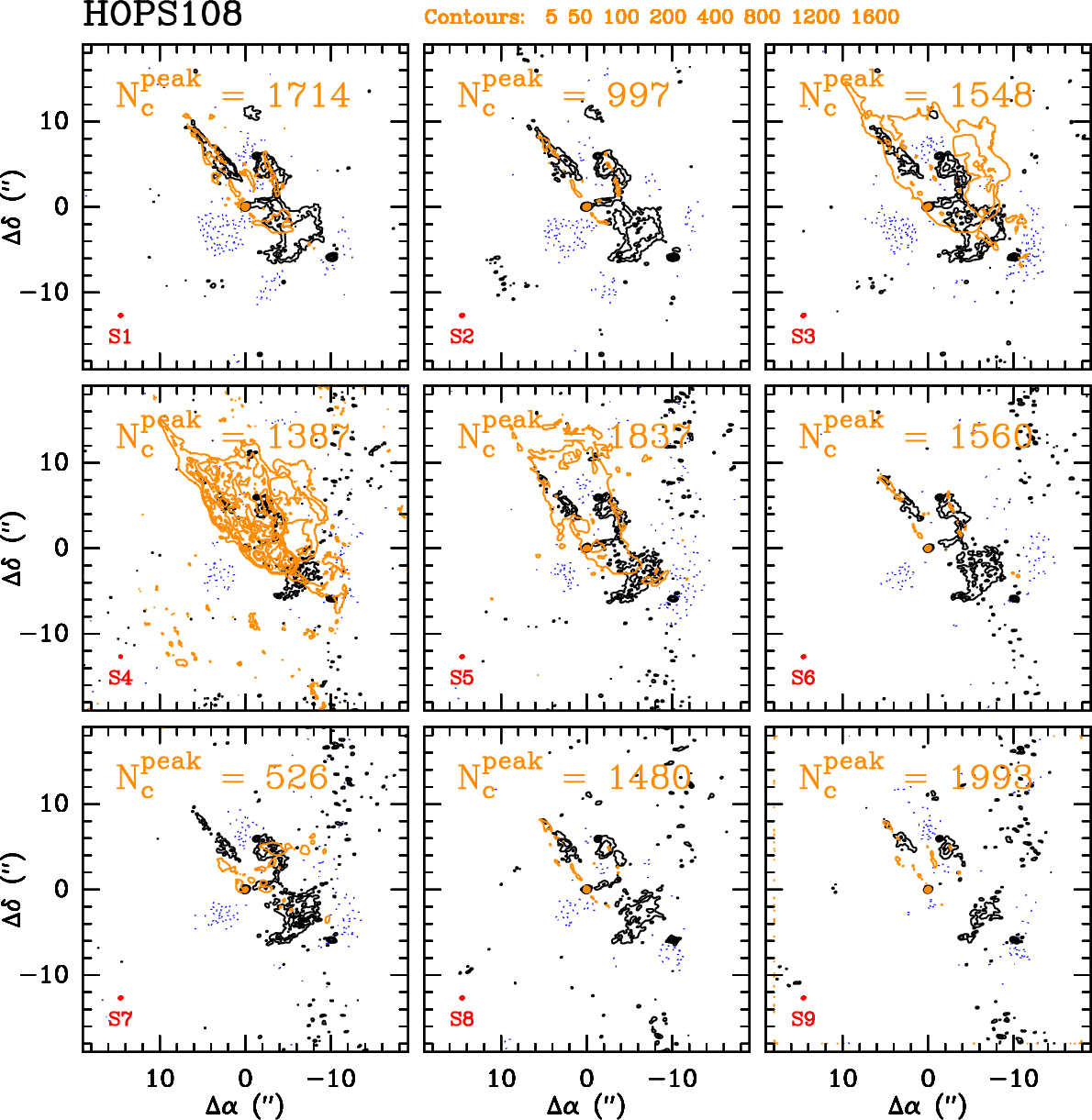}}}
\caption{Same as Fig.~\ref{f:chancount_svs13a}, but for HOPS108.}
\label{f:chancount_hops108}
\end{figure}

\begin{figure}[!ht]
\centerline{\resizebox{\hsize}{!}{\includegraphics[angle=0]{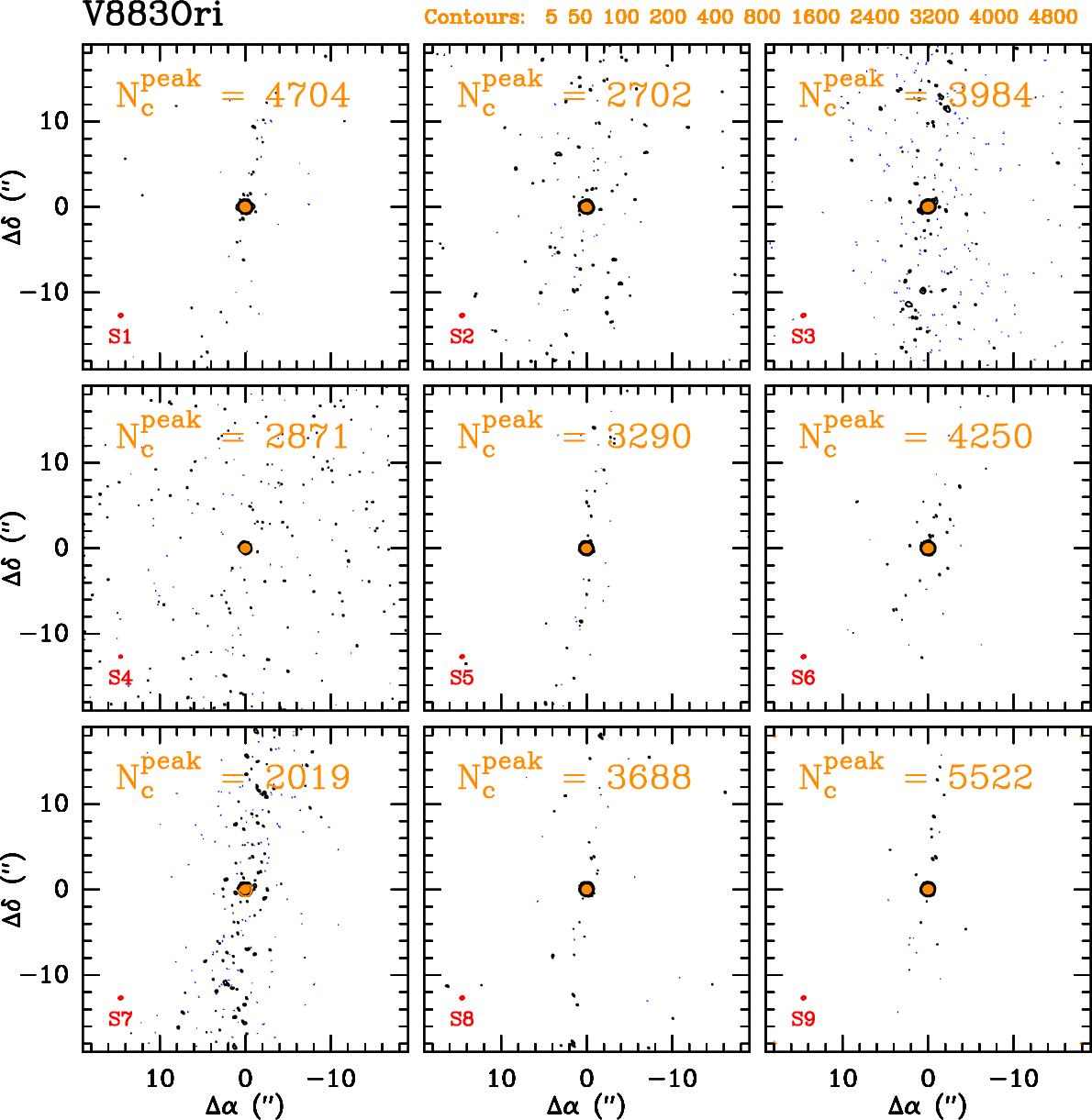}}}
\caption{Same as Fig.~\ref{f:chancount_svs13a}, but for V883Ori.}
\label{f:chancount_v883ori}
\end{figure}

\begin{figure}[!ht]
\centerline{\resizebox{\hsize}{!}{\includegraphics[angle=0]{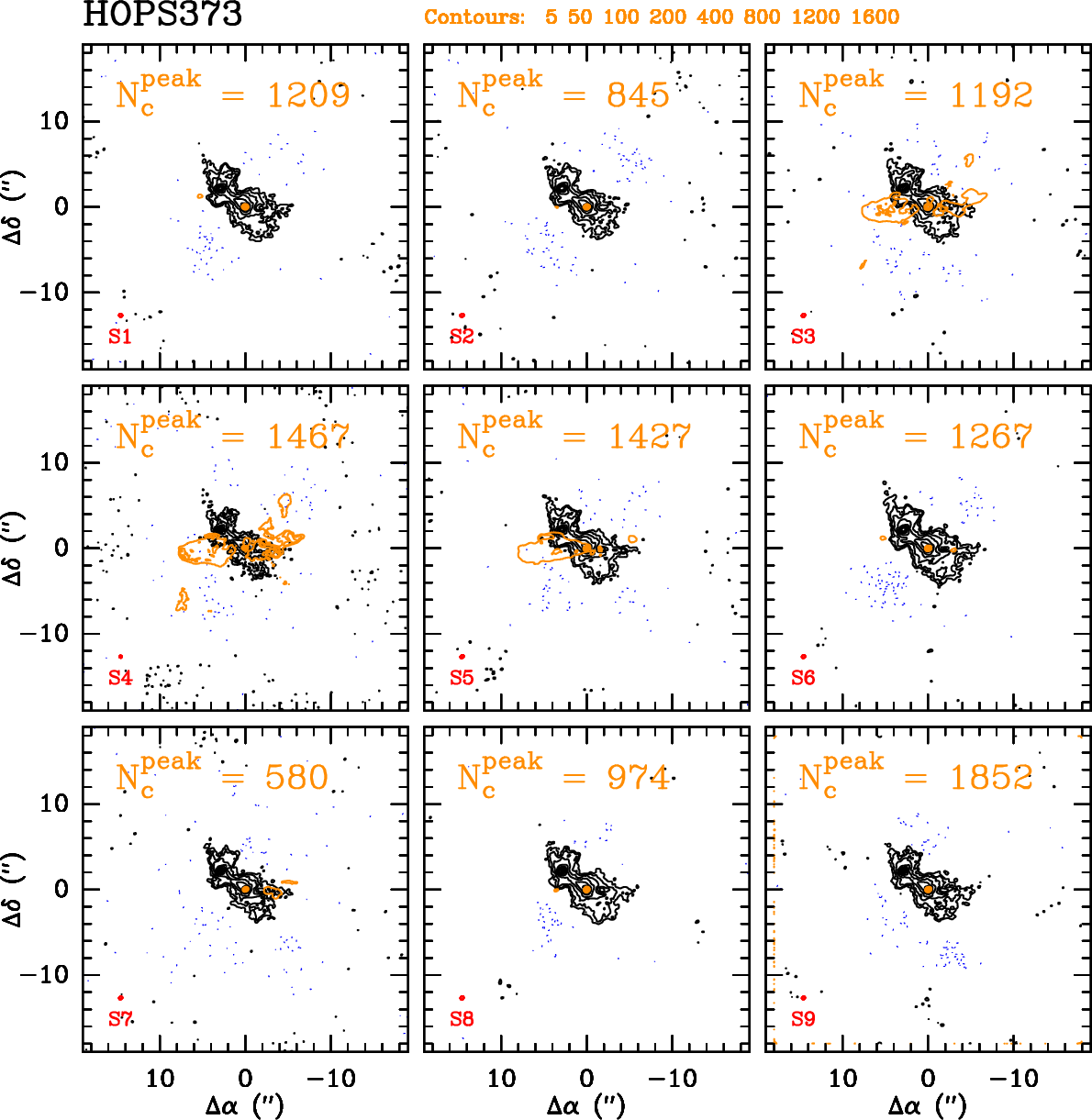}}}
\caption{Same as Fig.~\ref{f:chancount_svs13a}, but for HOPS373.}
\label{f:chancount_hops373}
\end{figure}

\begin{figure}[!ht]
\centerline{\resizebox{\hsize}{!}{\includegraphics[angle=0]{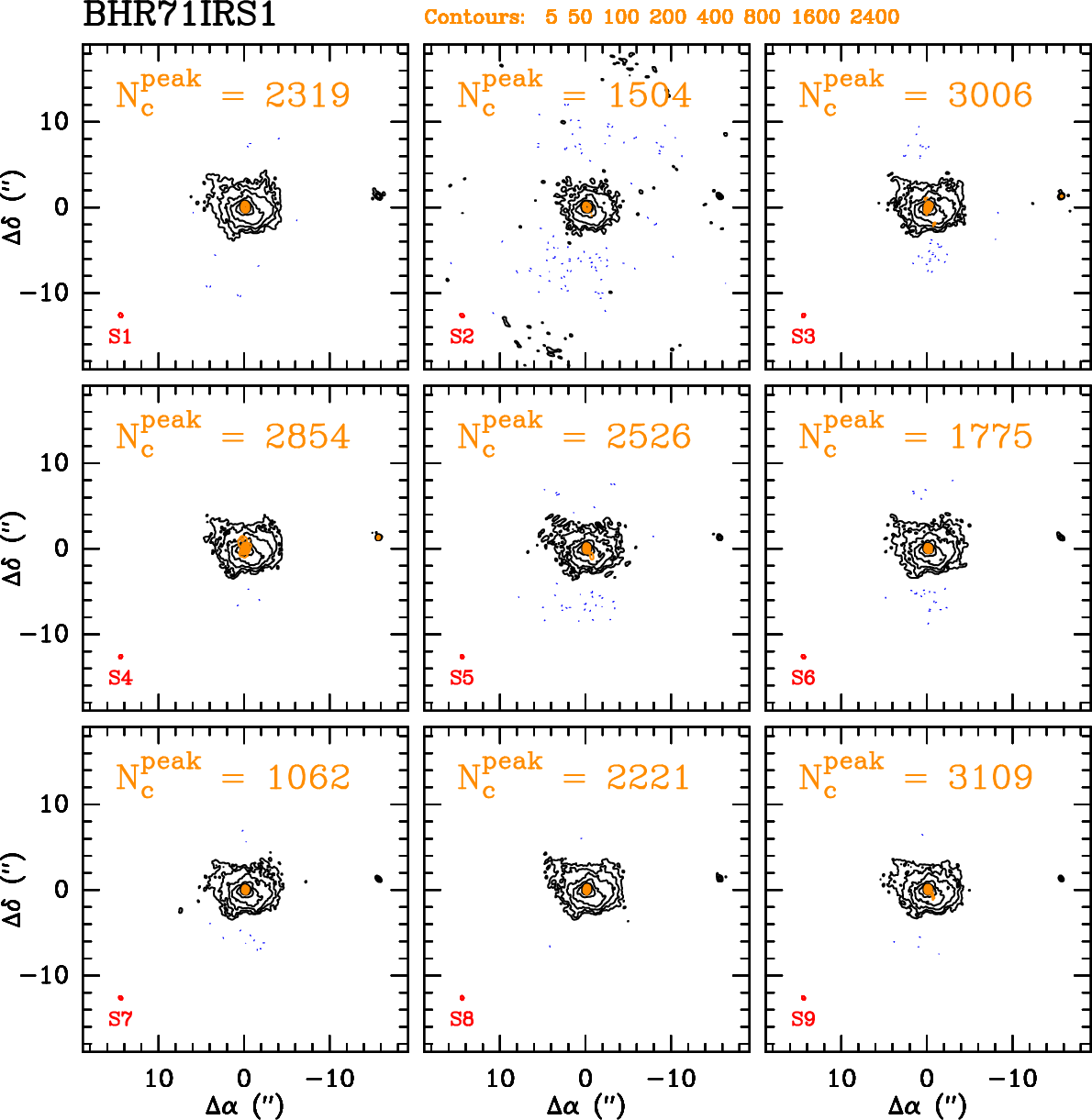}}}
\caption{Same as Fig.~\ref{f:chancount_svs13a}, but for BHR71IRS1.}
\label{f:chancount_bhr71}
\end{figure}

\begin{figure}[!ht]
\centerline{\resizebox{\hsize}{!}{\includegraphics[angle=0]{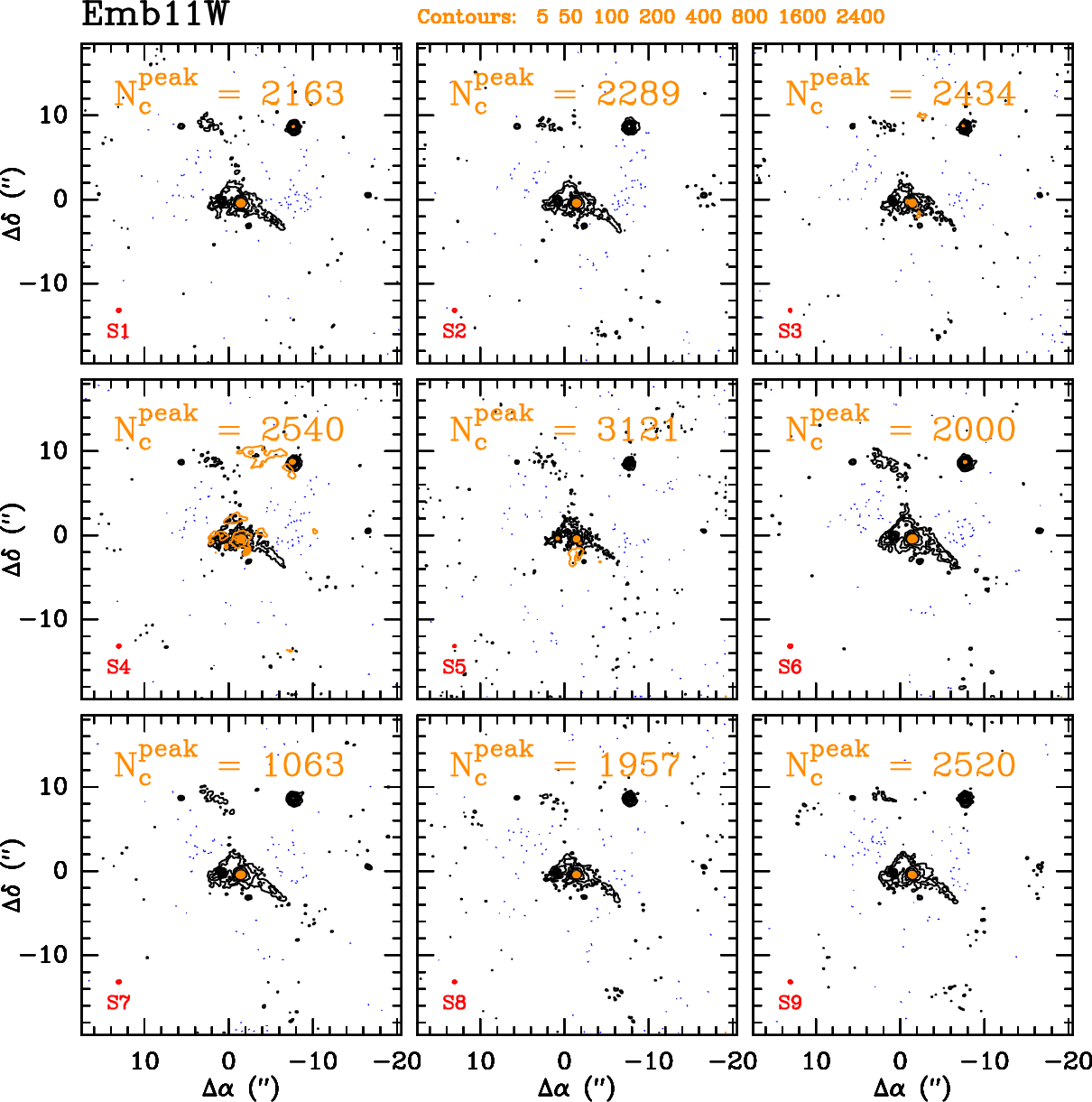}}}
\caption{Same as Fig.~\ref{f:chancount_svs13a}, but for Emb11W.}
\label{f:chancount_emb11w}
\end{figure}

\begin{figure}[!ht]
\centerline{\resizebox{\hsize}{!}{\includegraphics[angle=0]{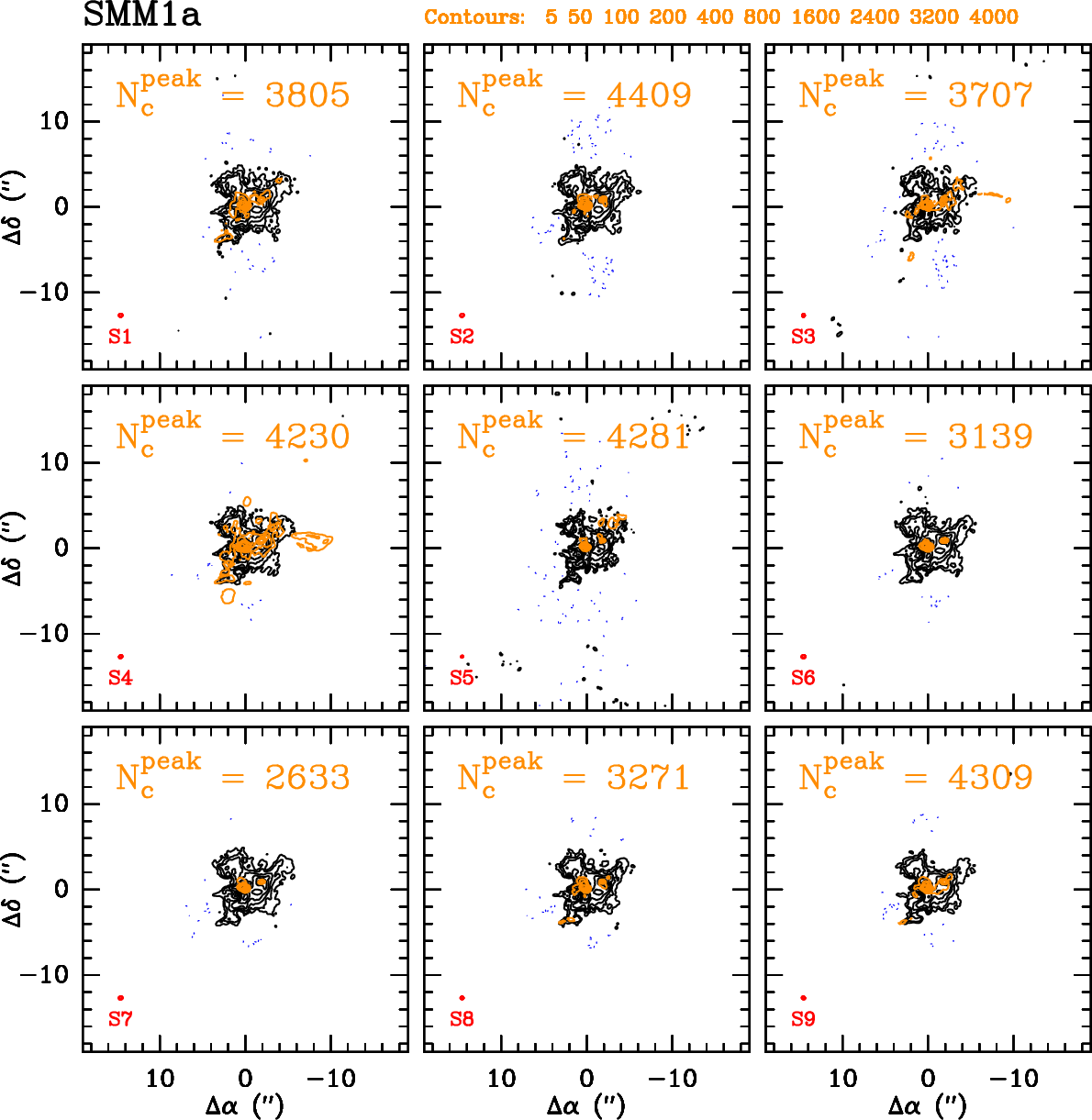}}}
\caption{Same as Fig.~\ref{f:chancount_svs13a}, but for SMM1a.
}
\label{f:chancount_smm1a}
\end{figure}

\begin{figure}[!ht]
\centerline{\resizebox{\hsize}{!}{\includegraphics[angle=0]{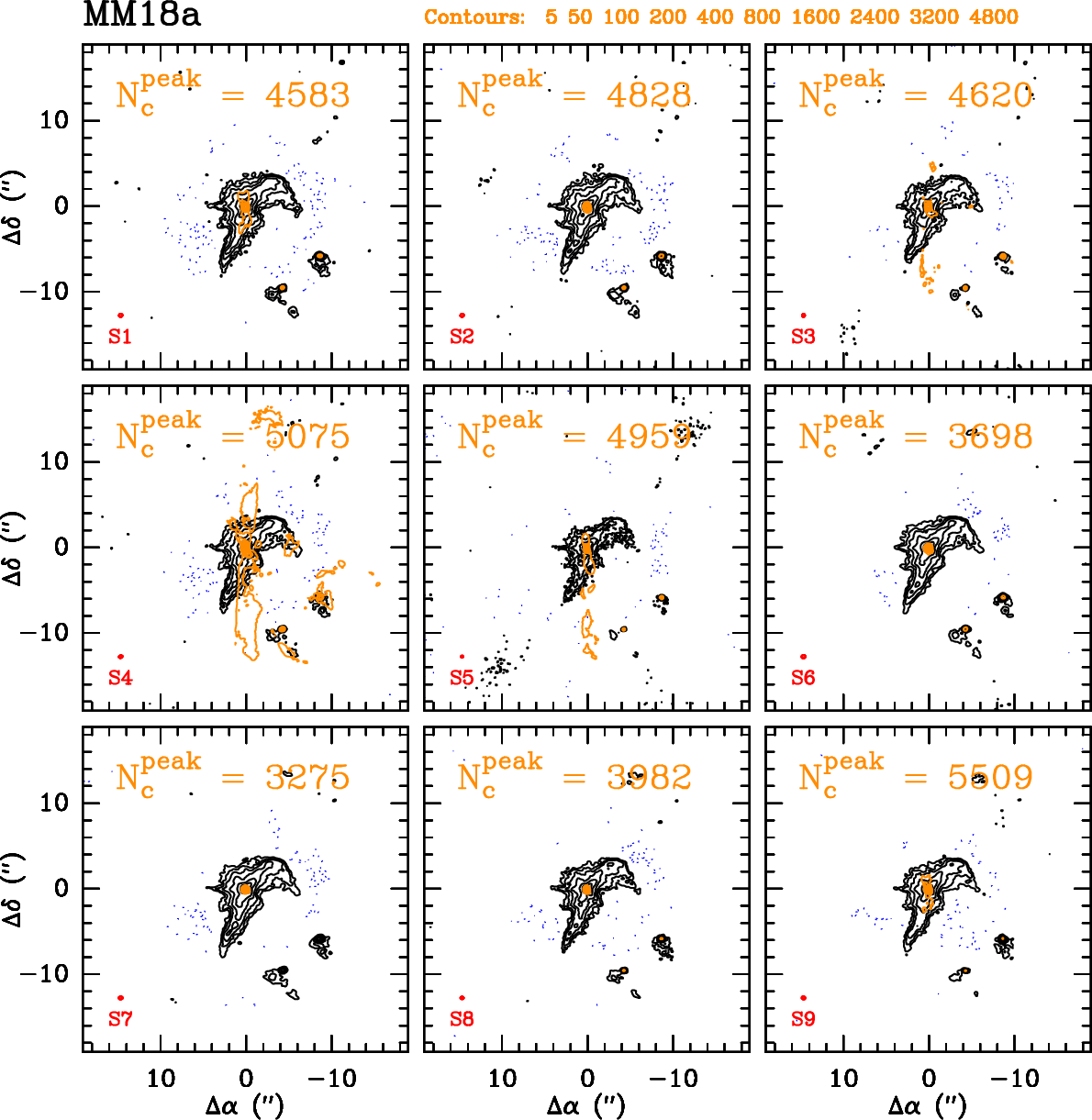}}}
\caption{Same as Fig.~\ref{f:chancount_svs13a}, but for MM18a.
}
\label{f:chancount_mm18a}
\end{figure}

\begin{figure}[!ht]
\centerline{\resizebox{\hsize}{!}{\includegraphics[angle=0]{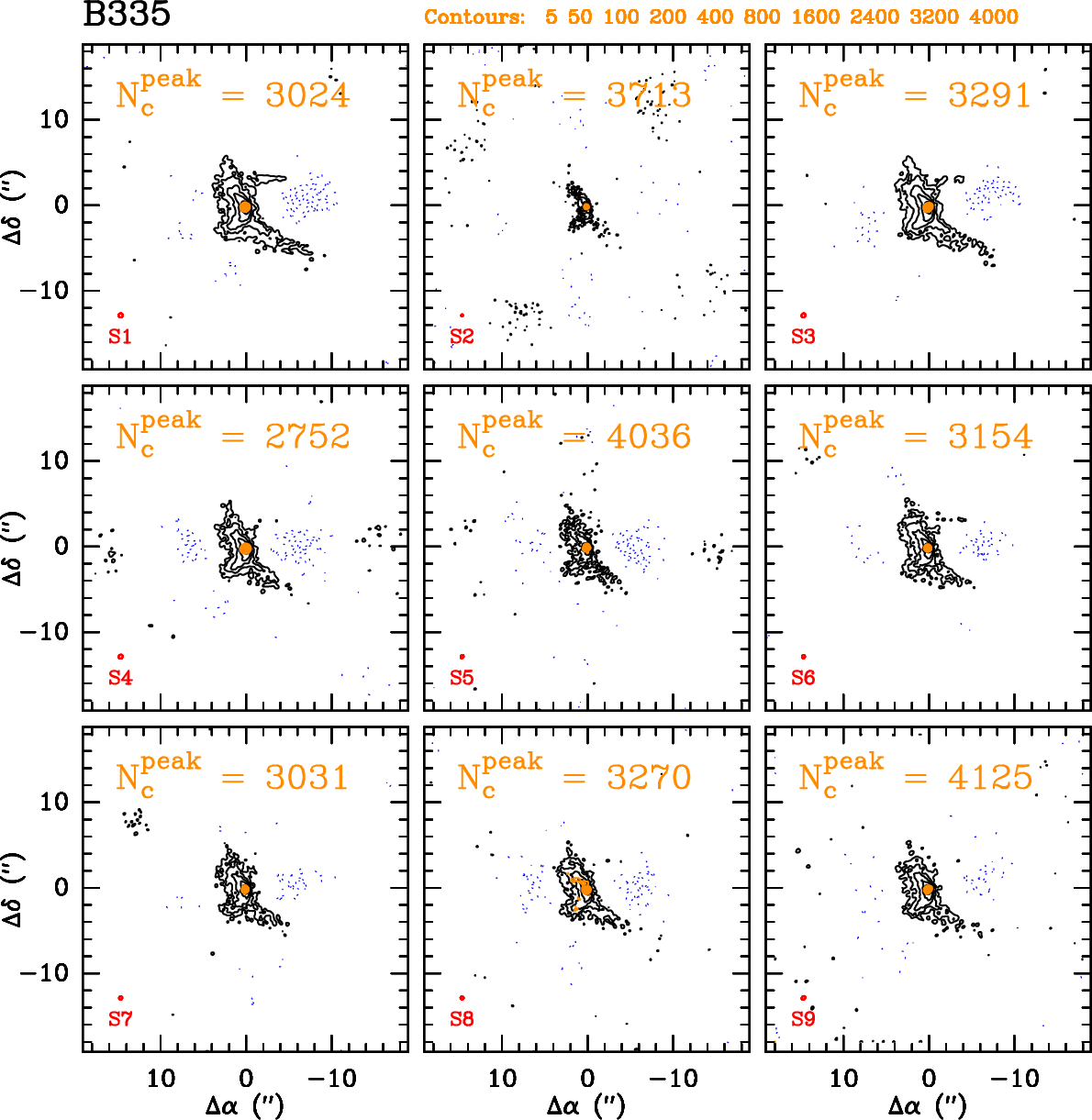}}}
\caption{Same as Fig.~\ref{f:chancount_svs13a}, but for B335.}
\label{f:chancount_b335}
\end{figure}

We computed a channel count map for each spectral window of each setup of each
field of view separately. For display purposes, we added the counts obtained 
for the four spectral windows of each given setup. The resulting channel count 
maps for each setup are displayed with orange contours in 
Figs.~\ref{f:chancount_svs13a}--\ref{f:chancount_b335} for the 11 fields 
covered by COMPASS. At least one compact region with a high channel count peak 
is detected in each field. A few maps also reveal bipolar structures that are 
associated with outflows. Most of the counts tracing outflow structures come 
from intermediate-energy transitions of methanol. For each
individual map of each spectral window of each setup, we fitted two-dimensional 
Gaussian functions to each region that showed compact channel count contours.
The coordinates of the median peak positions of these compact hot corino 
sources are listed in Table~\ref{t:chancountpos}. This table also provides an
estimate of the systemic velocity of each source. This velocity corresponds to
the best-fit velocity obtained in a preliminary LTE model of the emission of 
$^{13}$CH$_3$OH. In the cases where several velocity components were detected,
we provide the velocity of the dominating component or an unweighted mean of 
the dominating components.

\begin{table}[!ht]
 \begin{center}
 \caption{
 Frequency ranges masked before counting the channels.
}
\label{t:freqmasks}
 \vspace*{-1.2ex}
 \begin{tabular}{crll}
 \hline\hline
\multicolumn{1}{c}{Frequency mask\tablefootmark{a}} & \multicolumn{1}{c}{$\Delta f$\tablefootmark{b}} & \multicolumn{1}{c}{Molecule} & \multicolumn{1}{c}{Sp. window\tablefootmark{c}} \\ 
 \hline
 279068 -- 279082 & 14 & $^{34}$SO$_2$ &  S1W25 \\ 
 279423 -- 279437 & 14 & $^{34}$SO$_2$ &  S1W25 \\ 
 279505 -- 279519 & 14 & N$_2$H$^+$ &  S1W25 \\ 
 279550 -- 279564 & 14 & HDCS &  S1W25 \\ 
 279671 -- 279695 & 24 & OCS &  S1W25 \\ 
 280790 -- 280824 & 34 & SO$_2$ &  S1W27 \\ 
 281508 -- 281539 & 31 & H$_2$CO &  S1W29 \\ 
 281677 -- 281701 & 24 & SO$_2$ &  S1W29 \\ 
 281742 -- 281785 & 43 & SO$_2$ &  S1W29 \\ 
 282020 -- 282054 & 34 & SO$_2$ &  S1W31 \\ 
 282105 -- 282125 & 20 & NH$_2$D &  S1W31 \\ 
 282281 -- 282305 & 24 & SO$_2$ &  S1W31 \\ 
 282374 -- 282388 & 14 & c-C$_3$H$_2$ &  S1W31 \\ 
 282913 -- 282927 & 14 & D$^{13}$CO$^+$ &  S2W25 \\ 
 283172 -- 283196 & 24 & $^{34}$SO$_2$ &  S2W25 \\ 
 283218 -- 283232 & 14 & D$_2$$^{13}$CO &  S2W25 \\ 
 283435 -- 283449 & 14 & H$_2$$^{13}$CO &  S2W25 \\ 
 283441 -- 283456 & 15 & H$_2$C$^{17}$O &  S2W25 \\ 
 283448 -- 283486 & 38 & SO$_2$ &  S2W25 \\ 
 283986 -- 284000 & 14 & H$_2$$^{13}$CO &  S2W27 \\ 
 283993 -- 284008 & 15 & H$_2$C$^{17}$O &  S2W27 \\ 
 284110 -- 284128 & 18 & H$_2$$^{13}$CO &  S2W27 \\ 
 284119 -- 284138 & 19 & H$_2$C$^{17}$O &  S2W27 \\ 
 284270 -- 284284 & 14 & HDCS &  S2W27 \\ 
 284625 -- 284639 & 14 & H$_2$$^{13}$CO &  S2W29 \\ 
 284633 -- 284653 & 20 & H$_2$C$^{17}$O &  S2W29 \\ 
 284680 -- 284694 & 14 & D$^{13}$CN &  S2W29 \\ 
 284700 -- 284719 & 19 & OC$^{34}$S &  S2W29 \\ 
 284798 -- 284812 & 14 & c-C$_3$H$_2$ &  S2W29 \\ 
 284906 -- 284920 & 14 & c-C$_3$H$_2$ &  S2W29 \\ 
 284991 -- 285006 & 15 & c-C$_3$H$_2$ &  S2W29 \\ 
 285165 -- 285184 & 19 & $^{18}$OCS &  S2W29 \\ 
 285732 -- 285761 & 29 & SO$_2$ &  S2W31 \\ 
 285789 -- 285803 & 14 & c-C$_3$H$_2$ &  S2W31 \\ 
 286050 -- 286336 & 286 & Atm. O$_3$ & S2W31 S3W25 \\ 
 286136 -- 286150 & 14 & artefact? &  S2W31 \\ 
 286287 -- 286301 & 14 & H$_2$C$^{18}$O &  S2W31 \\ 
 286333 -- 286347 & 14 & SO &  S3W25 \\ 
 286404 -- 286428 & 24 & SO$_2$ &  S3W25 \\ 
 286498 -- 286512 & 14 & $^{33}$SO &  S3W25 \\ 
 286800 -- 286815 & 15 & NH$_2$D &  S3W25 \\ 
 287210 -- 287229 & 19 & D$_2$$^{13}$CO & S3W25 S3W27 \\ 
 287472 -- 287498 & 26 & D$_2$CO &  S3W27 \\ 
 287657 -- 287683 & 26 & CH$_3$OH &  S3W27 \\ 
 288137 -- 288151 & 14 & DCO$^+$ & S3W27 S3W29 \\ 
 288159 -- 288178 & 19 & OC$^{33}$S &  S3W29 \\ 
 288492 -- 288506 & 14 & CCD &  S3W29 \\ 
 288503 -- 288537 & 34 & SO$_2$ &  S3W29 \\ 
 288547 -- 288562 & 15 & CCD &  S3W29 \\ 
 288907 -- 289056 & 149 & Atm. O$_3$ & S3W29 S3W31 \\ 
 289195 -- 289221 & 26 & C$^{34}$S &  S3W31 \\ 
 289631 -- 289657 & 26 & DCN &  S3W31 \\ 
 290097 -- 290123 & 26 & CH$_3$OH &  S4W25 \\ 
 290555 -- 290574 & 19 & $^{34}$SO &  S4W25 \\ 
 290609 -- 290635 & 26 & H$_2$CO &  S4W25 \\ 
 290894 -- 290913 & 19 & O$^{13}$CS &  S4W27 \\ 
 291224 -- 291250 & 26 & H$_2$CO &  S4W27 \\ 
 291366 -- 291396 & 30 & H$_2$CO &  S4W27 \\ 
 291479 -- 291498 & 19 & C$^{33}$S &  S4W27 \\ 
\hline
\end{tabular}
 \end{center}
 \vspace*{-2.5ex}
\end{table}
\addtocounter{table}{-1}
\begin{table}[!ht]
 \begin{center}
 \caption{
 continued.
}
\label{t:freqmasks}
 \vspace*{-1.2ex}
 \begin{tabular}{crll}
 \hline\hline
\multicolumn{1}{c}{Frequency mask\tablefootmark{a}} & \multicolumn{1}{c}{$\Delta f$\tablefootmark{b}} & \multicolumn{1}{c}{Molecule} & \multicolumn{1}{c}{Sp. window\tablefootmark{c}} \\ 
 \hline
 291737 -- 291751 & 14 & D$_2$$^{13}$CO &  S4W27 \\ 
 291739 -- 291753 & 14 & D$_2$CO &  S4W27 \\ 
 291826 -- 291852 & 26 & OCS &  S4W29 \\ 
 291934 -- 291960 & 26 & H$_2$CO &  S4W29 \\ 
 292759 -- 292774 & 15 & D$_2$CO &  S4W31 \\ 
 292779 -- 292793 & 14 & H$_2$$^{13}$CS &  S4W31 \\ 
 292876 -- 292890 & 14 & SO$_2$ &  S4W31 \\ 
 293042 -- 293056 & 14 & D$_2$CO &  S4W31 \\ 
 293113 -- 293134 & 21 & H$_2$$^{13}$CO &  S4W31 \\ 
 293126 -- 293140 & 14 & H$_2$C$^{17}$O &  S4W31 \\ 
 293174 -- 293188 & 14 & D$_2$CO &  S4W31 \\ 
 293452 -- 293472 & 20 & $^{33}$SO &  S4W31 \\ 
 293895 -- 293929 & 34 & CS &  S5W25 \\ 
 294360 -- 294375 & 15 & N$^{34}$S &  S5W25 \\ 
 294750 -- 294766 & 16 & N$^{34}$S &  S5W27 \\ 
 295349 -- 295363 & 14 & SO &  S5W27 \\ 
 295389 -- 295408 & 19 & $^{34}$SO & S5W27 S5W29 \\ 
 295417 -- 295438 & 21 & H$_2$$^{34}$S &  S5W29 \\ 
 296157 -- 296181 & 24 & SO$_2$ &  S5W29 \\ 
 296418 -- 296432 & 14 & NH$_2$D &  S5W31 \\ 
 296528 -- 296542 & 14 & SO$_2$ &  S5W31 \\ 
 296480 -- 296670 & 190 & SO &  S5W31 \\ 
 296557 -- 296571 & 14 & OC$^{34}$S &  S5W31 \\ 
 296566 -- 296580 & 14 & $^{18}$OCS &  S5W31 \\ 
 296400 -- 296660 & 260 & $^{30}$SiO &  S5W31 \\ 
 296570 -- 296584 & 14 & D$_2$CO &  S5W31 \\ 
 297771 -- 297795 & 24 & SO$_2$ &  S6W25 \\ 
 297867 -- 297900 & 33 & H$_2$$^{33}$S &  S6W25 \\ 
 298239 -- 298270 & 31 & $^{33}$SO &  S6W27 \\ 
 298251 -- 298265 & 14 & $^{34}$SO &  S6W27 \\ 
 298559 -- 298593 & 34 & SO$_2$ &  S6W27 \\ 
 298683 -- 298697 & 14 & HCS$^+$ &  S6W27 \\ 
 299196 -- 299210 & 14 & H$_2$C$^{34}$S &  S6W29 \\ 
 299300 -- 299334 & 34 & SO$_2$ &  S6W29 \\ 
 299690 -- 299713 & 23 & NS &  S6W29 \\ 
 299945 -- 300205 & 260 & $^{29}$SiO &  S6W31 \\ 
 300087 -- 300111 & 24 & NS &  S6W31 \\ 
 300159 -- 300173 & 14 & OC$^{33}$S &  S6W31 \\ 
 300185 -- 300199 & 14 & c-C$_3$H$_2$ &  S6W31 \\ 
 300261 -- 300285 & 24 & SO$_2$ &  S6W31 \\ 
 300492 -- 300523 & 31 & H$_2$S &  S6W31 \\ 
 300820 -- 300849 & 29 & H$_2$CO & S6W31 S7W25 \\ 
 301067 -- 301084 & 17 & $^{33}$SO &  S7W25 \\ 
 301216 -- 301406 & 190 & SO &  S7W25 \\ 
 301479 -- 301493 & 14 & H$_2$$^{13}$CS &  S7W25 \\ 
 301730 -- 301950 & 220 & Atm. O$_3$ & S7W25 S7W27 \\ 
 301885 -- 301909 & 24 & SO$_2$ &  S7W27 \\ 
 302229 -- 302243 & 14 & SO$_2$ &  S7W27 \\ 
 302356 -- 302382 & 26 & CH$_3$OH &  S7W27 \\ 
 302703 -- 302717 & 14 & HD$^{13}$CO &  S7W29 \\ 
 302900 -- 302926 & 26 & CH$_3$OH &  S7W29 \\ 
 303008 -- 303022 & 14 & O$^{13}$CS &  S7W29 \\ 
 303353 -- 303379 & 26 & CH$_3$OH &  S7W29 \\ 
 303497 -- 303511 & 14 & H$_2$C$^{34}$S &  S7W29 \\ 
 303627 -- 303641 & 14 & HDCS &  S7W31 \\ 
 303746 -- 303760 & 14 & H$_2$C$^{34}$S &  S7W31 \\ 
 303795 -- 303809 & 14 & H$_2$C$^{34}$S &  S7W31 \\ 
 303750 -- 304010 & 260 & SiO &  S7W31 \\ 
 303986 -- 304000 & 14 & OCS &  S7W31 \\ 
 304007 -- 304197 & 190 & SO &  S7W31 \\ 
\hline
\end{tabular}
 \end{center}
 \vspace*{-2.5ex}
\end{table}
\addtocounter{table}{-1}
\begin{table}[!ht]
 \begin{center}
 \caption{
 continued.
}
\label{t:freqmasks}
 \vspace*{-1.2ex}
 \begin{tabular}{crll}
 \hline\hline
\multicolumn{1}{c}{Frequency mask\tablefootmark{a}} & \multicolumn{1}{c}{$\Delta f$\tablefootmark{b}} & \multicolumn{1}{c}{Molecule} & \multicolumn{1}{c}{Sp. window\tablefootmark{c}} \\ 
 \hline
 304194 -- 304220 & 26 & CH$_3$OH &  S7W31 \\ 
 304301 -- 304320 & 19 & H$_2$CS &  S7W31 \\ 
 305200 -- 305214 & 14 & DNC &  S8W25 \\ 
 305459 -- 305485 & 26 & CH$_3$OH &  S8W27 \\ 
 305716 -- 305737 & 21 & NH$_2$D &  S8W27 \\ 
 306384 -- 306410 & 26 & D$_2$CO &  S8W29 \\ 
 306732 -- 306746 & 14 & SO$_2$ &  S8W29 \\ 
 307152 -- 307178 & 26 & CH$_3$OH &  S8W29 \\ 
 307178 -- 307197 & 19 & SO$_2$ & S8W29 S8W31 \\ 
 307966 -- 307980 & 14 & $^{18}$OCS &  S8W31 \\ 
 308288 -- 308302 & 14 & H$_2$C$^{34}$S &  S9W25 \\ 
 308411 -- 308425 & 14 & HDCO &  S9W25 \\ 
 308413 -- 308427 & 14 & OC$^{34}$S &  S9W25 \\ 
 308415 -- 308434 & 19 & N$_2$D$^+$ &  S9W25 \\ 
 308743 -- 308762 & 19 & H$_2$CS &  S9W25 \\ 
 308979 -- 308998 & 19 & HDCS &  S9W25 \\ 
 309276 -- 309302 & 26 & CH$_3$OH &  S9W27 \\ 
 309495 -- 309514 & 19 & SO &  S9W27 \\ 
 309780 -- 309794 & 14 & HDCS &  S9W27 \\ 
 309928 -- 309947 & 19 & HDCS &  S9W29 \\ 
 310024 -- 310038 & 14 & HDCS &  S9W29 \\ 
 310038 -- 310052 & 14 & HDCS &  S9W29 \\ 
 310179 -- 310205 & 26 & CH$_3$OH &  S9W29 \\ 
 310754 -- 310768 & 14 & HDCS &  S9W29 \\ 
 311478 -- 311494 & 16 & $^{34}$SO$_2$ &  S9W31 \\ 
\hline
\end{tabular}
 \end{center}
 \vspace*{-2.5ex}
 \tablefoot{
\tablefoottext{a}{Frequency range of the mask in MHz.}
 \tablefoottext{b}{Frequency width of the mask in MHz.}
 \tablefoottext{c}{SxWyz means spectral window yz of setup x.}
}
 \end{table}
 
\section{Overview of full spectra extracted for the peak position for BHR71IRS1}
\begin{figure*}[!htb]
\resizebox{0.95\hsize}{!}{\includegraphics{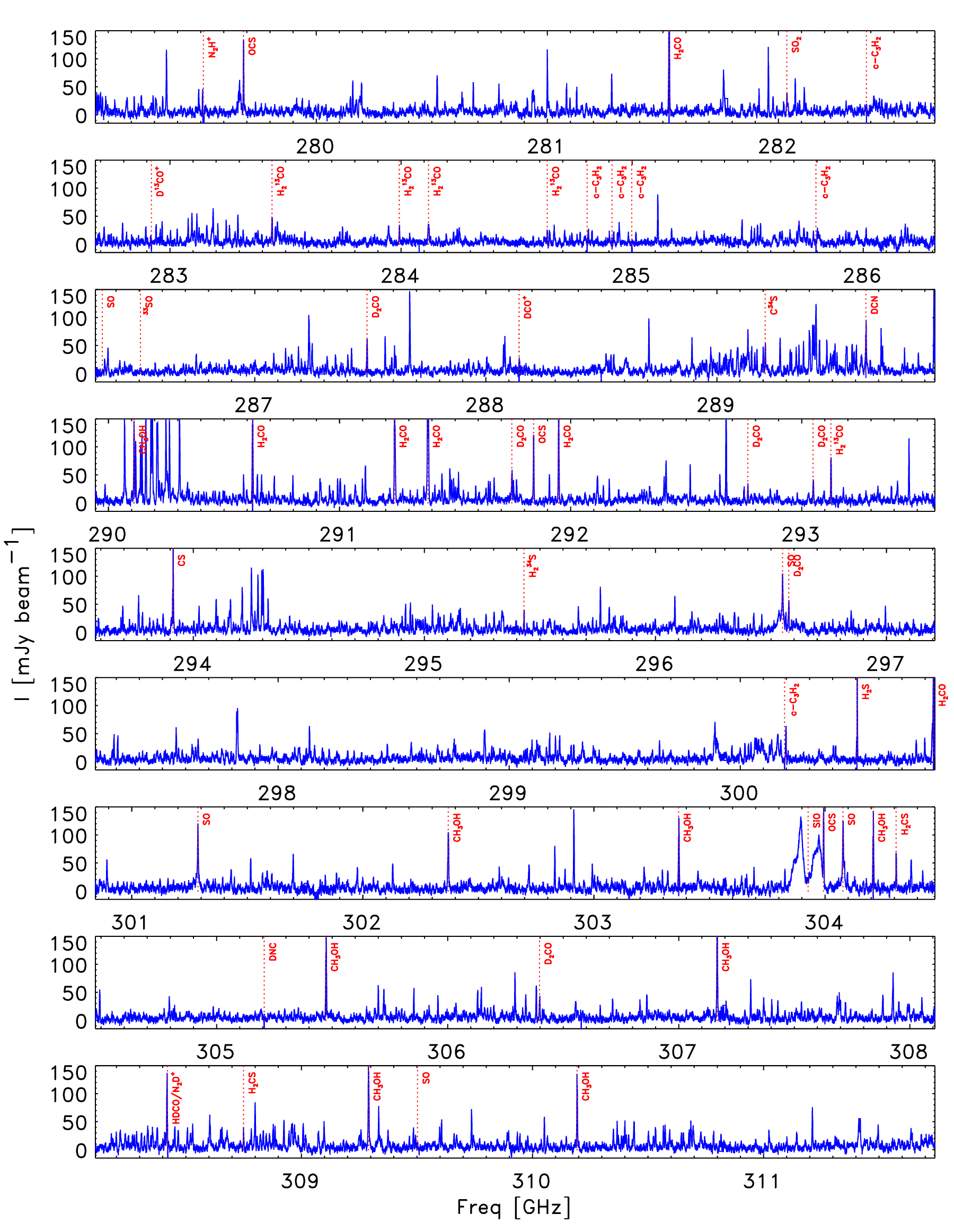}}
\caption{Spectrum over the full frequency range covered by the survey extracted toward the continuum peak position in BHR71-IRS1. The nine rows represent (approximately) the individual tunings or scheduling blocks that were used to cover this range. Selected major species are indicated with vertical dotted lines and labels.}\label{bhr71fullspec}
\end{figure*}

\begin{figure*}[!htb]
\resizebox{\hsize}{!}{\includegraphics{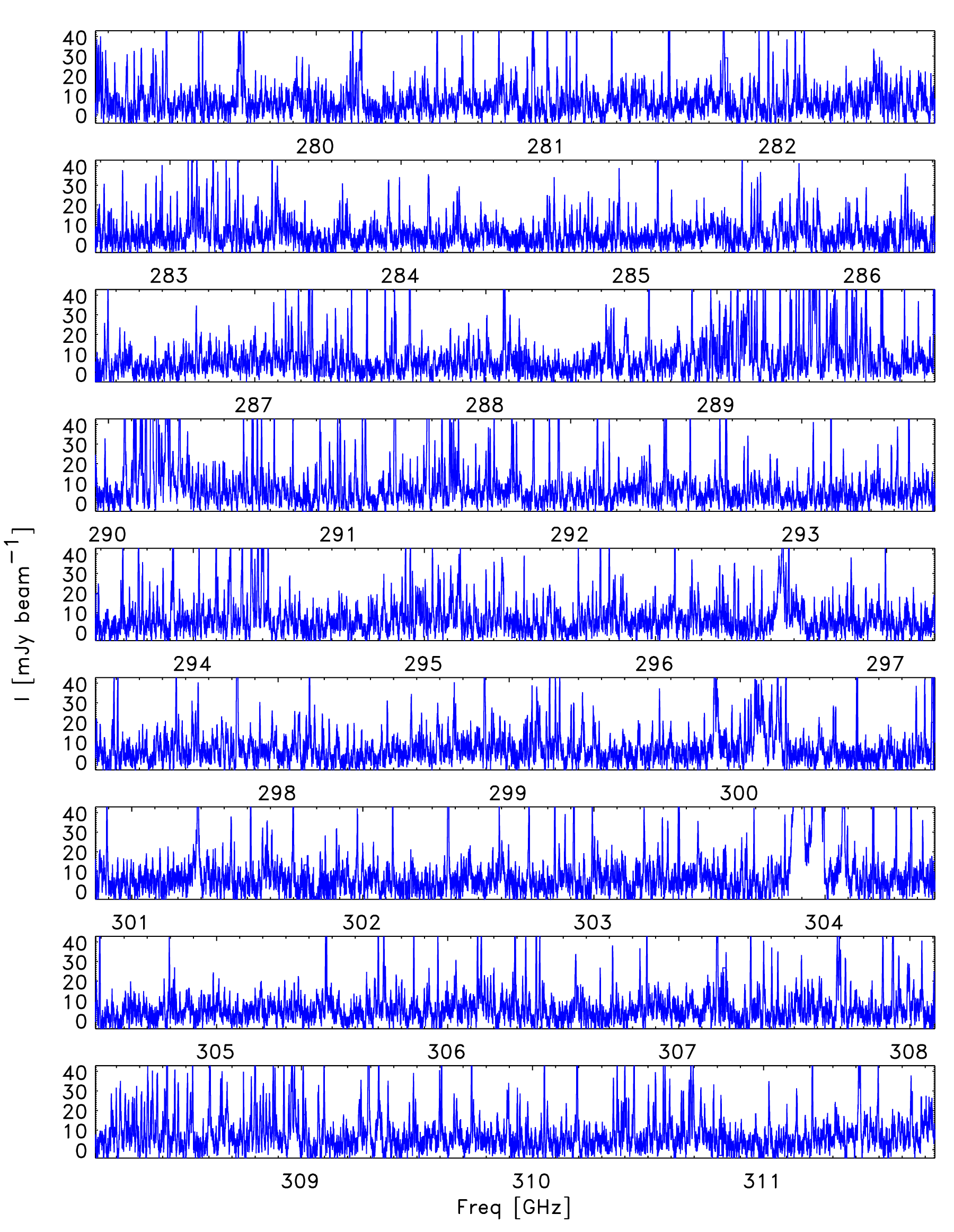}}
\caption{As in Fig.~\ref{bhr71fullspec}, but zoomed in on a range of the Y-axis corresponding roughly to emission from -3$\sigma$ to 30$\sigma$ in integrated intensity.}\label{bhr71fullspec_zoom}
\end{figure*}

Figures~\ref{bhr71fullspec} and \ref{bhr71fullspec_zoom} of this section shows the full spectra extracted for the continuum peak position of BHR71IRS1 (see Sect.~\ref{overviewspectra}).

\end{document}